\documentclass{aa}
\usepackage{graphicx, graphics}
\usepackage{CJK}
\usepackage{bbm}
\usepackage{amsmath, amssymb, bm}
\usepackage[nodayofweek]{datetime}
\input{prepcitationformat.txt}
\usepackage{txfonts}
\usepackage{ulem}

\usepackage{tikz}

\definecolor{lime}{HTML}{A6CE39}
\DeclareRobustCommand{\orcidicon}{%
    \raisebox{-3pt}{\begin{tikzpicture}
    \filldraw [lime, yshift=-2pt] (0, 0) circle [radius=0.16]
    node[white] {\raisebox{1pt}{\hspace{0.5pt}\fontfamily{qag}\selectfont\tiny i\scalebox{0.8}{D}}};
    \end{tikzpicture}}
    \hspace{-2.5mm}
    \vspace{-0.25pt}
}

\global\def\tablenotemark#1{{\color{blue}{\normalfont\textsuperscript{\scriptsize #1}}}} % changed by Bin Ren to \scriptsize to match published tablenotemark

\newdateformat{monthyearday}{%
  \THEYEAR\ \monthname[\THEMONTH] \twodigit{\THEDAY}}
\newdateformat{daymonthyear}{%
  \twodigit{\THEDAY}\ \monthname[\THEMONTH] \THEYEAR}
\newcommand\numberthis{\addtocounter{equation}{1}\tag{\theequation}}

\newcommand{\orcidauthor}[2]{#2\href{http://orcid.org/#1}{\orcidicon}}

\titlerunning{\textit{HST} Debris Disk Color}
\authorrunning{Ren et al.}

\begin{document}
\begin{CJK*}{UTF8}{gbsn}
\title{Debris Disk Color with the \textit{Hubble Space Telescope}}

\author{
\orcidauthor{0000-0003-1698-9696}{Bin B. Ren (任彬)}\inst{\ref{inst-oca}, \ref{inst-uga}, \ref{inst-cit}}
\and
\orcidauthor{0000-0002-4388-6417}{Isabel Rebollido}\inst{\ref{inst-stsci}} 
\and
\orcidauthor{0000-0002-9173-0740}{\'Elodie Choquet}\inst{\ref{inst-lam}}
\and
\orcidauthor{0000-0003-4229-8936}{Wen-Han Zhou (周文翰)}\inst{\ref{inst-oca}}
\and
\orcidauthor{0000-0002-3191-8151}{Marshall D. Perrin}\inst{\ref{inst-stsci}}
\and
\orcidauthor{0000-0002-4511-5966}{Glenn Schneider}\inst{\ref{inst-ua}}
\and
\orcidauthor{0000-0001-9325-2511}{Julien Milli}\inst{\ref{inst-uga}}
\and
\orcidauthor{0000-0002-9977-8255}{Schuyler G. Wolff}\inst{\ref{inst-ua}}
\and
\orcidauthor{0000-0002-8382-0447}{Christine H. Chen}\inst{\ref{inst-stsci}}
\and
\orcidauthor{0000-0002-1783-8817}{John H. Debes}\inst{\ref{inst-stsci}}
\and
J. Brendan Hagan\inst{\ref{inst-stsci}}
\and
\orcidauthor{0000-0003-4653-6161}{Dean C. Hines}\inst{\ref{inst-stsci}}
\and
\orcidauthor{0000-0001-6205-9233}{Maxwell A. Millar-Blanchaer}\inst{\ref{inst-ucsb}}
\and
Laurent Pueyo\inst{\ref{inst-stsci}}
\and
\orcidauthor{0000-0002-2989-3725}{Aki Roberge}\inst{\ref{inst-gsfc}}
\and
Eugene Serabyn\inst{\ref{inst-jpl}}
\and
\orcidauthor{0000-0003-2753-2819}{R\'emi Soummer}\inst{\ref{inst-stsci}}
}

\institute{
Universit\'{e} C\^{o}te d'Azur, Observatoire de la C\^{o}te d'Azur, CNRS, Laboratoire Lagrange, F-06304 Nice, France; \email{\url{bin.ren@oca.eu}} \label{inst-oca}
\and
Universit\'{e} Grenoble Alpes, CNRS, Institut de Plan\'{e}tologie et d'Astrophysique (IPAG), F-38000 Grenoble, France \label{inst-uga}
\and
Department of Astronomy, California Institute of Technology, MC 249-17, 1200 East California Boulevard, Pasadena, CA 91125, USA \label{inst-cit}
\and
Space Telescope Science Institute (STScI), 3700 San Martin Drive, Baltimore, MD 21218, USA \label{inst-stsci}
\and
Aix Marseille Univ, CNRS, CNES, LAM, Marseille, France \label{inst-lam}
\and
Steward Observatory, The University of Arizona, Tucson, AZ 85721, USA \label{inst-ua}
\and
Jet Propulsion Laboratory, California Institute of Technology, 4800 Oak Grove Drive, Pasadena, CA 91109, USA \label{inst-jpl}
\and
Department of Physics, University of California, Santa Barbara, CA 93106, USA \label{inst-ucsb}
\and
Astrophysics Science Division, NASA Goddard Space Flight Center, Greenbelt, MD 20771, USA \label{inst-gsfc}
}

\date{Received 14 November 2022; revised 7 February 2023; accepted 8 February 2023}

\abstract
{Multi-wavelength scattered light imaging of debris disks may inform dust properties including typical size and mineral composition. Existing studies have investigated a small set of individual systems across a variety of imaging instruments and filters, calling for uniform comparison studies to systematically investigate dust properties.}
{We obtain the surface brightness of dust particles in debris disks by post-processing coronagraphic imaging observations, and compare the multi-wavelength reflectance of dust. For a sample of resolved debris disks, we perform a systematic analysis on the reflectance properties of their birth rings.}
{We reduced the visible and near-infrared images of $23$ debris disk systems hosted by A through M stars using two coronagraphs onboard the \textit{Hubble Space Telescope}: the STIS instrument observations centering at $0.58~\mu$m, and the NICMOS instrument at $1.12~\mu$m or $1.60~\mu$m. For proper recovery of debris disks, we used classical reference differential imaging for STIS, and adopted non-negative matrix factorization with forward modeling for NICMOS. By dividing disk signals by stellar signals to take into account of intrinsic stellar color effects, we systematically obtained and compared the reflectance of debris birth rings at ${\approx}90^\circ$ scattering angle.}
{Debris birth rings typically exhibit a blue color at ${\approx}90^\circ$ scattering angle. As the stellar luminosity increases, the color tends to be more neutral. A likely L-shaped color-albedo distribution indicates a clustering of scatterer properties.}
{The observed color trend correlates with the expected blow-out size of dust particles. The color--albedo clustering likely suggests different populations of dust in these systems. More detailed radiative transfer models with realistic dust morphology will contribute to explaining the observed color and color-albedo distribution of debris systems.}


\maketitle

\section{Introduction}
Debris disks are extrasolar analogs of the Asteroid Belt and the Kuiper Belt \citep[e.g.][]{hughes18}. They are composed of second-generation dust, in the sense that their life time is shorter than the age of their host star \citep[e.g.,][]{wyatt08}, and they are produced from and continuously replenished by collisional cascade of larger solid bodies \citep{dohnanyi69}. While collisional cascade produces small dust particles, radiation pressure can surpass gravity for small dust particles and blow certain dust particles out of stellar systems \citep[e.g.,][]{strubbe06, krivov06}. The balance of forces for dust particles results in a blow-out size, which ranges from sub-micron to several microns depending on both stellar properties and dust properties \citep[e.g., spectral type, dust composition, dust porosity:][]{arnold19}. Observationally, depending on the blow-out size and other dust properties in disks, there could exist noticeable differences \citep[e.g., scattering phase function:][]{munoz21}.

In the birth ring of a debris disk, dust particles under collisional cascade have an expected number distribution $n(a)\propto a^{-3.5}$ where $a$ is the particle size \citep[e.g.,][]{pan12}. Under the fact that the cross section of each particle is proportional to $a^2$, collisional cascade can make the smaller particles dominate more surface area of a debris disk. In reality, stellar radiation pressure can drive smaller particles to higher-eccentricity or even unbound orbits, resulting into blow-out sizes above which dust particles are bound. Nevertheless, the balance between radiation and gravity predicts that dust particles can be unbound only within a certain size range \citep[e.g.,][]{thebault19}, and that there is no stellar-radiation-driven blow-out size for certain later-type stars (e.g., M stars: \citealp{arnold19}). Other mechanisms, including stellar winds in M stars (e.g., AU Mic: \citealp{augereau06}, TWA~7: \citealp{olofsson20}), can also remove dust particles from stellar environments, complicating the size distribution of dust in debris systems. The joint effect of these mechanisms could lead to observational complexity for debris disks.

Studies on spectral energy distribution (SED) of debris disks showed that the ratio between dust temperature and blackbody temperature at the disk radius decreases with increasing stellar luminosity \citep[e.g.][]{pawellek14}. Although this trend can be explained by the hypothesis that typical dust size increases with stellar luminosity \citep{pawellek14, pawellek15}, the blackbody location of disks can offset from their resolved locations by a factor of ${\sim}4$ \citep[e.g., scattered light imaging:][]{esposito20}, since debris dust particles are inefficient emitters at longer wavelengths. This offset makes SED modeling a degenerate problem between dust property and disk location. With resolved disk images in scattered light, we can break these known degeneracies for the smallest dust in debris systems.

Using a variety of coronagraphic imaging instruments from both the ground (e.g., NaCo: \citealp{lagrange03, lenzen03}, GPI: \citealp{macintosh08}, SPHERE: \citealp{beuzit08}) and the space (e.g., ACS: \citealp{ford98}, NICMOS: \citealp{ramberg93}, STIS: \citealp{woodgate98}), multi-wavelength scattered light imaging studies  revealed dust properties for debris disks individually, such as 49~Ceti \citep{choquet17, pawellek19_49ceti}, AU~Mic \citep{fitzgerald07}, Beta~Pic \citep{golimowski06}, HD~15115 \citep{kalas07}, HD~32297 \citep{kalas05, duchene20}, HD~35841 \citep{esposito18}, HD~107146 \citep{ertel11}, HD~191089 \citep{ren19}, HD~192758 \citep{choquet18}, HR~4796A \citep{debes08, milli15, rodigas15, chen20, arriaga20}, and TWA~7 \citep{ren21}. These multi-wavelength studies, when further augmented with the advantage of uniform imaging exploration from identical instruments \citep[e.g., GPI debris disk survey:][]{esposito20}, would minimize the offsets from different instruments to enable uniform systematic studies of dust properties, thus bring forth essential information on the ensemble properties of debris disks in scattered light.

With debris disks resolved in scattered light, existing studies have investigated their ensemble properties, especially on scattering phase functions (SPFs) which depict the scattered light intensity dependence on scattering angles. \citet{hughes18} suggested that the SPFs of debris disks could follow a uniform trend, yet more recent observations with high-precision measurements showed diverse SPFs in different systems \citep[e.g.,][]{ren19, engler22} or even potential SPF change in different wavelengths \citep[e.g.,][]{ren20di}. In addition, SPF measurements could be impacted by instrumentation effects including convolution, by data reduction artifacts such as over-fitting and self-subtraction in high-contrast total intensity imaging, and by vertical thickness effects \citep[e.g.,][]{milli17, olofsson20}, these complications make it a necessity to study debris disks from another complementary perspective -- multi-band imaging \citep[e.g.,][]{chen20, arriaga20} -- to depict their collective properties.

Onboard the \textit{Hubble Space Telescope} (\textit{HST}), the Space Telescope Imaging Spectrograph \citep[STIS:][]{woodgate98} and Near Infrared Camera and Multi-Object Spectrometer \citep[NICMOS:][]{thompson92} instruments can offer unparalleled stability and sensitivity in the coronagraphic imaging of circumstellar disks from visible light to near infrared wavelengths. In comparison with protoplanetary disks that are relatively bright and facile to be attempted from ground-based extreme-adaptive-optics-equipped systems in polarized light \citep[e.g.,][]{avenhaus18, laws20}, \textit{HST} coronagraphs can offer both stable stellar point spread function (PSF) and optimal sensitivity for faint target imaging. These advanced instruments provide the most effective method for imaging faint debris disks in total intensity \citep[e.g., STIS:][]{krist10, krist12, schneider18}. In addition, the fact that \textit{HST} operates in vacuum makes it not only straightforward to calibrate detector readouts to physical units \citep[e.g.,][]{nicmosbook09} than ground-based observations \citep[e.g.,][]{milli15}, but also sensitive to the faintest materials such as debris halos that are elusive from the ground \citep[e.g., halos:][]{schneider18, ren19}.

With the high-stability, high-sensitivity, and high-spatial-resolution offered by \textit{HST}, resolved scattered light imaging of debris disks can directly probe the spatial and surface brightness distributions for the smallest dust particles within \citep[e.g.,][]{schneider14, schneider18}. When imaged at multiple wavelengths, the color information of the scatterers can inform dust properties \citep[e.g., composition, porosity:][]{debes08}. In addition, resolved imaging of debris disks enabled by application of advanced statistical methods, especially when applied to archival observations and recovering the hidden debris disks \citep[e.g.,][]{soummer14, choquet14}, can allow the study of dust properties to an unprecedented degree \citep[e.g., albedo:][]{choquet18}. Combining the advantages of multi-wavelength images offered by \textit{HST} and disk recovery from advanced methods, here we perform a uniform recovery and study of resolved debris disks to investigate their ensemble properties. We describe the observation and the data reduction procedures to recover resolved disk images in Sect.~\ref{sec-obs}, analyze the data in Sect.~\ref{sec-ana}, discuss our findings in Sect.~\ref{sec-diss}, and conclude this study in Sect.~\ref{sec-diss}.

\section{Observation \& Data Reduction}\label{sec-obs}
\begin{figure}[htb!]
	\includegraphics[width=0.46\textwidth]{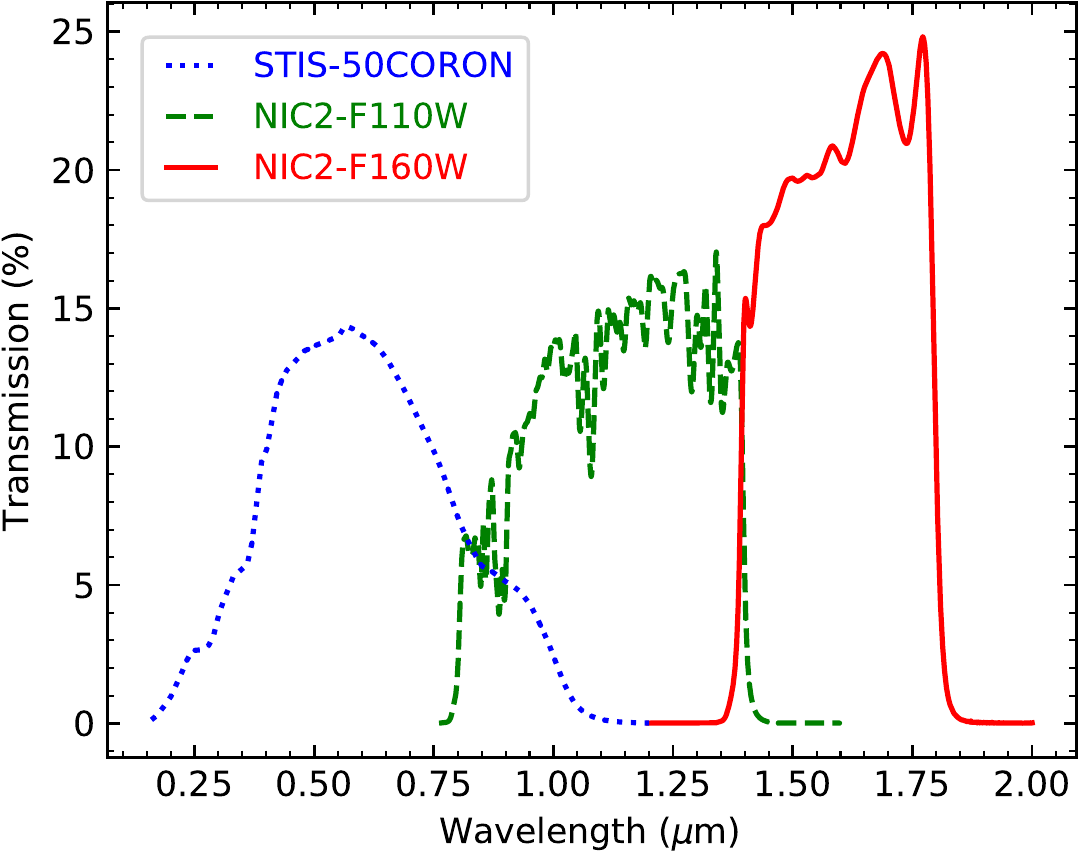}
    \caption{Transmission of the STIS-50CORON, NIC2-F110W, and NIC2-F160W coronagraphic filters used to image debris disks in this study.}
    \label{fig-transmission}
\end{figure}

We summarized a total of 23 systems observed in coronagraphic imaging mode using both STIS (filter: 50CORON; $\lambda_{\text{c}}=0.58\ \mu$m, pixel scale: $50.72$~mas~pixel$^{-1}$, \citealp{stisihb18}) and NICMOS Camera 2 (NIC2; filter: F110W or F160W; $\lambda_{\text{c}}=1.12\ \mu$m or $1.60\ \mu$m, pixel scale: $75.65$~mas~pixel$^{-1}$, \citealp{nicmosbook09}). In Fig.~\ref{fig-transmission}, we display the transmission curves of the three filters \citep[obtained from][]{Rodrigo12,Rodrigo20}. The debris systems are: 49~Ceti, AU~Mic, Beta~Pic, HD~377, HD~15115, HD~15745, HD~30447, HD~32297, HD~35650, HD~35841, HD~61005, HD~104860, HD~110058, HD~131835, HD~141569A, HD~141943, HD~181327, HD~191089, HD~192758, HD~202917, HR~4796A, TWA~7, and TWA~25. We summarize the properties\footnote{Unless otherwise specified, the error bars calculated in this paper are $1\sigma$.} of the targets in Table~\ref{tab:prop}, and the exposure information in Table~\ref{tab:expinfo}.

\begin{table*}
\centering
\setlength{\tabcolsep}{4.5pt}
\caption{Property of debris disk hosts observed by \textit{HST}/STIS and \textit{HST}/NICMOS \label{tab:prop}}
\begin{tabular}{c l c c c c c c c c c c}    \hline\hline
 id	&\multicolumn{1}{c}{Target} 	&SpType & $V$ &  Distance & $T_{\rm eff}$  & $M_{\rm star}$ & $L_{\rm star}$&   $\log g$& $a_{\rm BO}$  & \multicolumn{2}{c}{Reference}\\
 	&	&  & (mag) & (pc) &  (K)  & ($M_\sun$) & ($L_\sun$)& (cm s$^{-2}$) &($\mu$m) & SpType & $V$-mag\\
 (1)	& \multicolumn{1}{c}{(2)} & (3) & (4) &  (5)  & (6) & (7)& (8) &(9)& (10) & (11) & (12) \\ \hline
$a$ 	& 49~Ceti        	& A1V            	& 5.61 	& $57.23_{-0.18}^{+0.18}$ 	& $9000_{-400}^{+170}$ 	& $2.2_{-0.3}^{+0.3}$ 	& $15.7_{-0.7}^{+0.7}$ 	& $4.32_{-0.07}^{+0.07}$ 	& $2.5_{-0.4}^{+0.4}$ 	& 1 	& 13 \\
$b$ 	& AU~Mic         	& M1V            	& 8.63 	& $9.714_{-0.002}^{+0.002}$ 	& $3992_{-166}^{+150}$ 	& $0.710_{-0.014}^{+0.014}$ 	& $0.073_{-0.004}^{+0.004}$ 	& $4.6_{-0.06}^{+0.06}$ 	& $0.036_{-0.002}^{+0.002}$ 	& 2 	& 14 \\
$c$ 	& Beta~Pic       	& A6V            	& 3.86 	& $19.63_{-0.06}^{+0.06}$ 	& $7100_{-300}^{+300}$ 	& $1.9_{-0.2}^{+0.2}$\tablenotemark{a}	& $8.97_{-0.07}^{+0.07}$\tablenotemark{a}	& $4.4_{-0.3}^{+0.3}$\tablenotemark{a}	& $1.65_{-0.17}^{+0.17}$ 	& 3 	& 15 \\
$d$ 	& HD~377         	& G2V            	& 7.59 	& $38.40_{-0.04}^{+0.04}$ 	& $5871_{-40}^{+30}$ 	& $1.07_{-0.13}^{+0.13}$ 	& $1.16_{-0.03}^{+0.03}$ 	& $4.44_{-0.08}^{+0.08}$ 	& $0.38_{-0.05}^{+0.05}$ 	& 4 	& 15 \\
$e$ 	& HD~15115       	& F4IV           	& 6.80 	& $48.77_{-0.07}^{+0.07}$ 	& $6811_{-150}^{+150}$ 	& $1.4_{-0.2}^{+0.2}$ 	& $3.73_{-0.15}^{+0.15}$ 	& $4.31_{-0.08}^{+0.08}$ 	& $0.90_{-0.15}^{+0.15}$ 	& 5 	& 16 \\
$f$ 	& HD~15745       	& F0             	& 7.49 	& $71.73_{-0.12}^{+0.12}$ 	& $6840_{-140}^{+130}$ 	& $1.5_{-0.3}^{+0.3}$ 	& $4.21_{-0.17}^{+0.17}$ 	& $4.27_{-0.09}^{+0.09}$ 	& $1.0_{-0.18}^{+0.18}$ 	& 6 	& 16 \\
$g$ 	& HD~30447       	& F3V            	& 7.86 	& $80.31_{-0.14}^{+0.14}$ 	& $6709_{-150}^{+130}$ 	& $1.5_{-0.3}^{+0.3}$ 	& $3.73_{-0.14}^{+0.14}$ 	& $4.31_{-0.09}^{+0.09}$ 	& $0.89_{-0.16}^{+0.16}$ 	& 7 	& 16 \\
$h$ 	& HD~32297       	& A6V            	& 8.14 	& $129.7_{-0.5}^{+0.5}$ 	& $7980_{-80}^{+170}$ 	& $1.9_{-0.3}^{+0.3}$ 	& $8.5_{-0.4}^{+0.4}$ 	& $4.36_{-0.08}^{+0.08}$ 	& $1.5_{-0.3}^{+0.3}$ 	& 8 	& 16 \\
$i$ 	& HD~35650       	& K6V            	& 9.05 	& $17.461_{-0.005}^{+0.005}$ 	& $4175_{-60}^{+150}$ 	& $0.66_{-0.08}^{+0.08}$ 	& $0.129_{-0.009}^{+0.009}$ 	& $4.6_{-0.11}^{+0.11}$ 	& $0.069_{-0.010}^{+0.010}$ 	& 4 	& 17 \\
$j$ 	& HD~35841       	& F3V            	& 8.90 	& $103.07_{-0.14}^{+0.14}$ 	& $6305_{-80}^{+200}$ 	& $1.3_{-0.2}^{+0.2}$ 	& $2.43_{-0.10}^{+0.10}$ 	& $4.38_{-0.09}^{+0.09}$ 	& $0.64_{-0.11}^{+0.11}$ 	& 1 	& 16 \\
$k$ 	& HD~61005       	& G8V            	& 8.22 	& $36.45_{-0.02}^{+0.02}$ 	& $5507_{-120}^{+60}$ 	& $0.97_{-0.12}^{+0.12}$ 	& $0.636_{-0.014}^{+0.014}$ 	& $4.54_{-0.07}^{+0.07}$ 	& $0.23_{-0.03}^{+0.03}$ 	& 3 	& 16 \\
$l$ 	& HD~104860      	& G0/F9V         	& 7.91 	& $45.19_{-0.04}^{+0.04}$ 	& $5939_{-100}^{+60}$ 	& $1.11_{-0.15}^{+0.15}$ 	& $1.18_{-0.04}^{+0.04}$ 	& $4.48_{-0.08}^{+0.08}$ 	& $0.37_{-0.05}^{+0.05}$ 	& 9 	& 16 \\
$m$ 	& HD~110058      	& A0V            	& 7.97 	& $130.1_{-0.5}^{+0.5}$ 	& $8039_{-190}^{+900}$ 	& $2.0_{-0.3}^{+0.3}$ 	& $9.4_{-0.5}^{+0.5}$ 	& $4.4_{-0.08}^{+0.08}$ 	& $1.6_{-0.3}^{+0.3}$ 	& 10 	& 16 \\
$n$ 	& HD~131835      	& A2IV           	& 7.86 	& $129.7_{-0.5}^{+0.5}$ 	& $8266_{-300}^{+500}$ 	& $2.1_{-0.3}^{+0.3}$ 	& $10.9_{-0.9}^{+0.9}$ 	& $4.37_{-0.08}^{+0.08}$ 	& $1.8_{-0.3}^{+0.3}$ 	& 7 	& 16 \\
$o$ 	& HD~141569      	& A2V            	& 7.12 	& $111.6_{-0.4}^{+0.4}$ 	& $8439_{-700}^{+200}$ 	& $2.1_{-0.4}^{+0.4}$ 	& $15.3_{-4.0}^{+4.0}$ 	& $4.2_{-0.3}^{+0.3}$ 	& $2.5_{-0.8}^{+0.8}$ 	& 11 	& 16 \\
$p$ 	& HD~141943      	& G2             	& 7.97 	& $60.14_{-0.08}^{+0.08}$ 	& $5673_{-110}^{+100}$ 	& $1.09_{-0.14}^{+0.14}$ 	& $2.07_{-0.06}^{+0.06}$ 	& $4.22_{-0.08}^{+0.08}$ 	& $0.66_{-0.09}^{+0.09}$ 	& 4 	& 17 \\
$q$ 	& HD~181327      	& F6V            	& 7.04 	& $47.78_{-0.07}^{+0.07}$ 	& $6436_{-160}^{+40}$ 	& $1.3_{-0.2}^{+0.2}$ 	& $2.88_{-0.11}^{+0.11}$ 	& $4.3_{-0.08}^{+0.08}$ 	& $0.76_{-0.13}^{+0.13}$ 	& 4 	& 16 \\
$r$ 	& HD~191089      	& F5V            	& 7.18 	& $50.11_{-0.05}^{+0.05}$ 	& $6450_{-180}^{+50}$ 	& $1.3_{-0.2}^{+0.2}$ 	& $2.74_{-0.10}^{+0.10}$ 	& $4.33_{-0.09}^{+0.09}$ 	& $0.72_{-0.12}^{+0.12}$ 	& 7 	& 16 \\
$s$ 	& HD~192758      	& F0V            	& 7.03 	& $66.50_{-0.14}^{+0.14}$ 	& $7200_{-200}^{+160}$ 	& $1.6_{-0.3}^{+0.3}$ 	& $5.4_{-0.2}^{+0.2}$ 	& $4.26_{-0.09}^{+0.09}$ 	& $1.2_{-0.2}^{+0.2}$ 	& 10 	& 19 \\
$t$ 	& HD~202917      	& G7V            	& 8.67 	& $46.71_{-0.03}^{+0.03}$ 	& $5506_{-100}^{+70}$ 	& $0.98_{-0.12}^{+0.12}$ 	& $0.668_{-0.016}^{+0.016}$ 	& $4.53_{-0.08}^{+0.08}$ 	& $0.24_{-0.03}^{+0.03}$ 	& 4 	& 19 \\
$u$ 	& HR~4796A       	& A0V            	& 5.77 	& $70.8_{-0.2}^{+0.2}$ 	& $9670_{-500}^{+100}$ 	& $2.5_{-0.3}^{+0.3}$ 	& $24.7_{-1.1}^{+1.1}$ 	& $4.35_{-0.06}^{+0.06}$ 	& $3.5_{-0.5}^{+0.5}$ 	& 7 	& 16 \\
$v$ 	& TWA~7          	& M2V            	& 10.91 	& $34.10_{-0.03}^{+0.03}$ 	& $4018_{-170}^{+150}$ 	& $0.46_{-0.09}^{+0.09}$ 	& $0.115_{-0.019}^{+0.019}$ 	& $4.18_{-0.17}^{+0.17}$ 	& $0.09_{-0.02}^{+0.02}$ 	& 4 	& 16 \\
$w$ 	& TWA~25         	& M0.5           	& 11.16 	& $53.60_{-0.07}^{+0.07}$ 	& $4020_{-160}^{+200}$ 	& $0.60_{-0.08}^{+0.08}$ 	& $0.23_{-0.02}^{+0.02}$ 	& $4.17_{-0.13}^{+0.13}$ 	& $0.14_{-0.02}^{+0.02}$ 	& 12 	& 17 \\ \hline
\end{tabular}

\begin{flushleft}

{\tiny \textbf{Notes}: Column (1): letter identifiers of the targets in this paper. Column (2): target name. Column (3): spectral type from the literature (column 11). Column (4): $V$-band magnitude from literature studies in Column (12). Column (5): distance computed from \textit{Gaia} DR3 parallaxes \citep{gaiaDR3}. Column (6): effective temperature from \textit{Gaia} DR2 \citep{gaiadr2}. Values in Column (7) for star mass, Column (8) for stellar luminosity, and Column (9) for stellar surface gravity are from the \textit{Transiting Exoplanet Survey Satellite} input catalog 
\citep{tess_input_catalog}. Column (10): expected dust blowout size for non-porous amorphous olivine using Equation~\eqref{eq-abo}. While M stars do not have sufficient radiation pressure to blow small dust out \citep[e.g.,][]{arnold19}, we report the corresponding blow-out sizes for color-size correlation analysis in Sect.~\ref{sec-color-amin}.

$^a${For Beta Pic, the uncertainties of $M_{\rm star}$, $L_{\rm star}$, and $\log g$ are scaled from \citet{david15}, \citet{anders19}, and \citet{gaiadr2}, respectively. If the upper and lower uncertainties are different, the bigger one is adopted.}
}

{\tiny \textbf{References}: In Column (11) and Column (12), the references are for spectral type and $V$-mag, respectively: (1) \citet{1988MSS...C04....0H}; (2) \citet{1989ApJS...71..245K}; (3) \citet{2006AJ....132..161G}; (4) \citet{2006AA...460..695T}; (5) \citet{1974AJ.....79..682H}; (6) \citet{1993yCat.3135....0C}; (7) \citet{1982MSS...C03....0H}; (8) \citet{2014ApJ...783...21R}; (9) \citet{2016MNRAS.458.2307K}; (10) \citet{1978MSS...C02....0H}; (11) \citet{2017AJ....154...31G}; (12) \citet{2014ApJ...786...97H}; (13) \citet{2000AA...355L..27H}; (14) \citet{2012AcA....62...67K}; (15) \citet{2002yCat.2237....0D}; (16) \citet{simbad}; (17) \citet{2012yCat.1322....0Z}.
}
\end{flushleft}

\end{table*}

\subsection{STIS}

Using STIS, we observed 4 systems (HD~30447, HD~35841, HD~141943, and HD~191089) under \textit{HST} GO-13381\footnote{\url{https://www.stsci.edu/cgi-bin/get-proposal-info?id=13381&observatory=HST}} (PI: M.~Perrin),  9 systems (49~Ceti, HD~377, HD~35650, HD~104860, HD~110058, HD~131835, HD~192758, TWA~7, and TWA~25) under \textit{HST} GO-15218\footnote{\url{https://www.stsci.edu/cgi-bin/get-proposal-info?id=15218&observatory=HST}} (PI: \'E.~Choquet). From the MAST archive,\footnote{\url{https://archive.stsci.edu}} we retrieved 6 systems (AU~Mic, HD~15115, HD~15745, HD~32297, HD~61005, and HD~181327) from \textit{HST} GO-12228 (PI: G.~Schneider; \citealp{schneider14}), 2 systems (HD~202917 and HR~4796A) from \textit{HST} GO-13786 (PI: G.~Schneider; \citealp{schneider16, schneider18}). For Beta~Pic, we retrieved its observations from three programs: SM2/ERO-7125 (PI: S.~Heap;  \citealp{heap00}), \textit{HST} GO-12551 (PI: D.~Apai; \citealp{apai15}) and \textit{HST} GO-12923 (PI: A.~Gaspar; \citealp{schneider17}). For HD~141569A, from three programs: \textit{HST} GO-8624 (PI: A.~Weinberger), \textit{HST} GO-8674 (PI: A.-M.~Lagrange; \citealp{mouillet01}) and \textit{HST} GO-13786 (PI: G.~Schneider; \citealp{konishi16}). 

\begin{figure*}[htb!]
	\includegraphics[width=\textwidth]{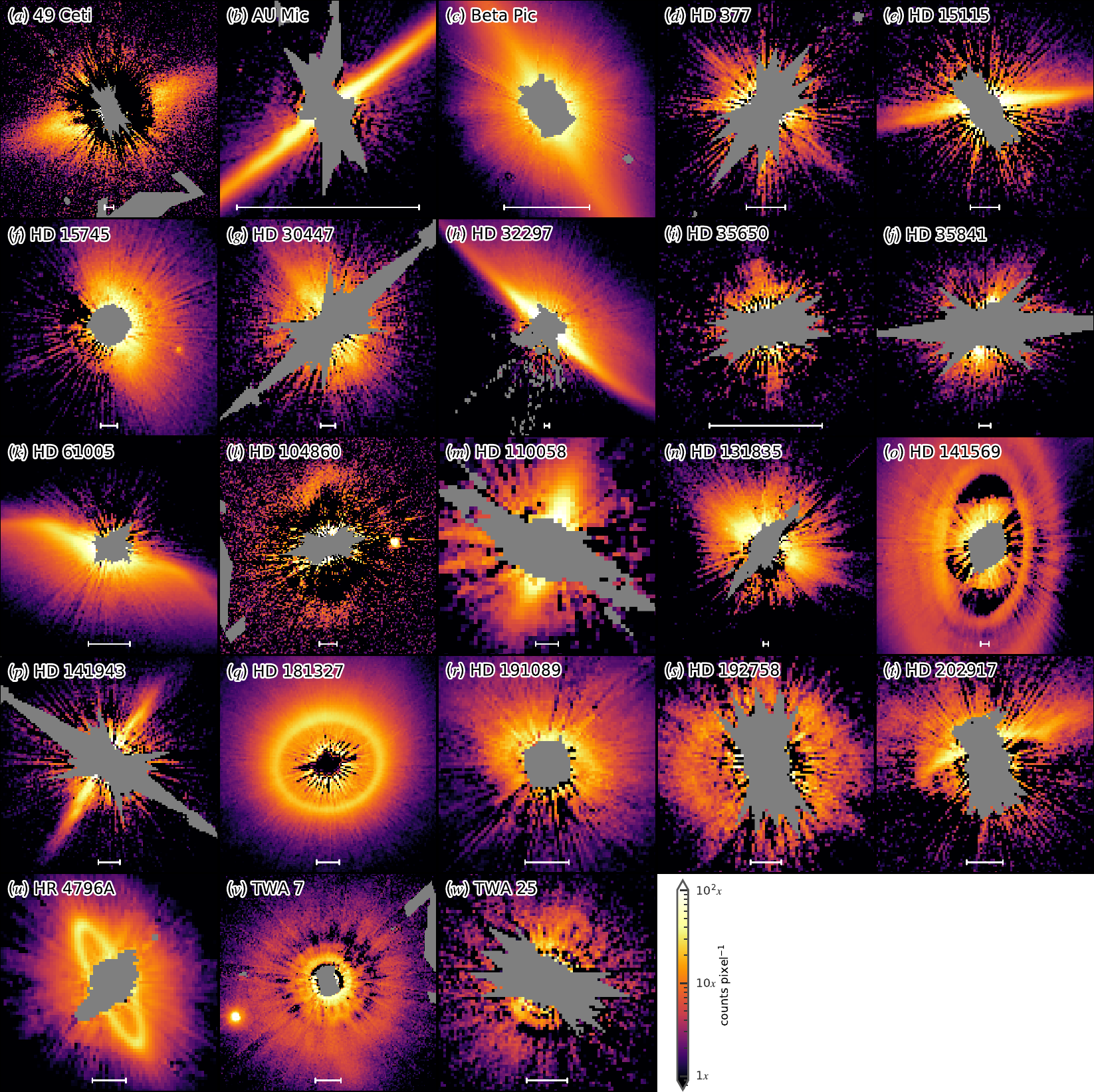}
    \caption{Surface brightness distribution of the STIS disks. The color bars are in log scale with arbitrary units to adjust for difference in disk surface brightness across our debris disk gallery, and the scale bars are 50~au.}
    \label{fig-stis}
\end{figure*}

For each target, we reduced the observation data with multi-roll combined PSF-template subtraction \citep[MRDI: ][]{schneider14} using its corresponding PSF reference images designated in each \textit{HST} program. We note that although HD~377 was previously observed in \textit{HST} GO-12291 (PI: J.~Krist), it was not recovered since the major axis of the disk coincides with either the STIS occulter or the diffraction spikes. In addition, we observed negligible difference between median-combined and mean-combined images, and thus we used the mean-combined MRDI images for a proper propagation of errors. We present the reduced images in Fig.~\ref{fig-stis}.

\subsection{NICMOS}
\begin{figure*}[htb!]
	\includegraphics[width=\textwidth]{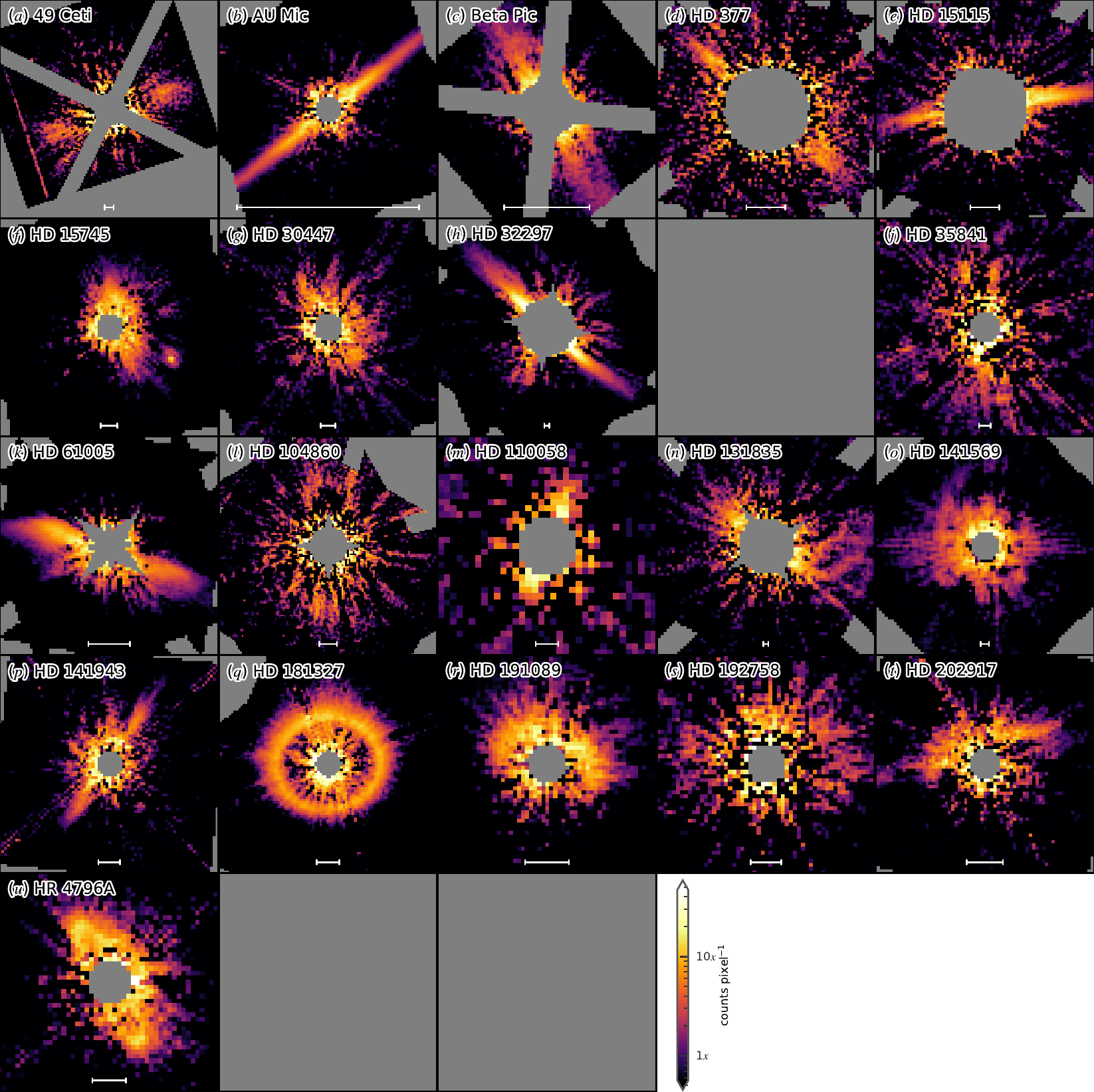}
    \caption{Surface brightness distribution of the NICMOS disks using the F110W filter. The color bars are in log scale with arbitrary units, and the scale bars are 50~au.}
    \label{fig-nicmos-110}
\end{figure*}

\begin{figure*}[htb!]
	\includegraphics[width=\textwidth]{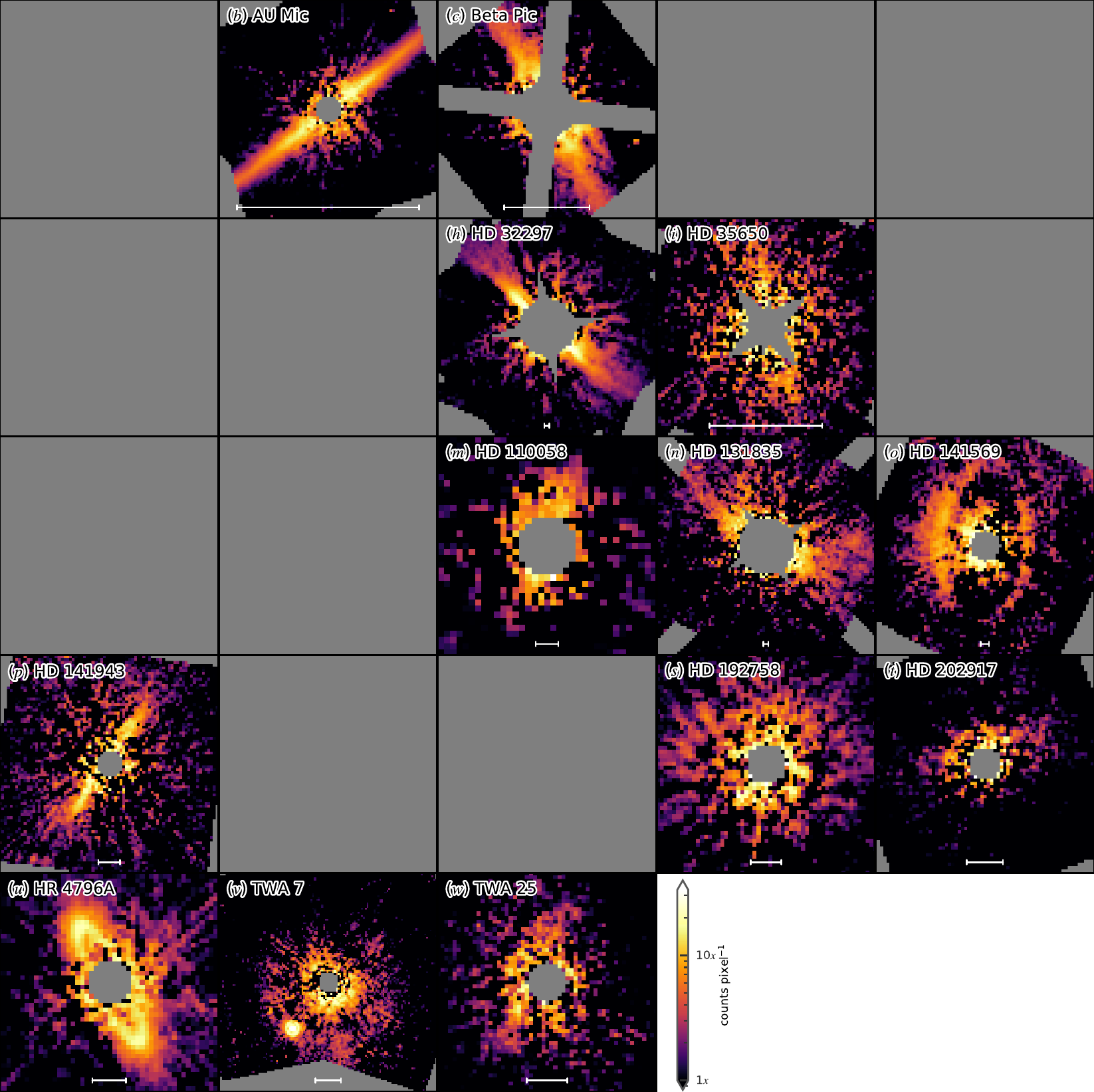}
    \caption{Surface brightness distribution of the NICMOS disks using the F160W filter. The color bars are in log scale with arbitrary units, and the scale bars are 50~au.}
    \label{fig-nicmos-160}
\end{figure*}

We assembled the NICMOS observations for the targets and their corresponding PSF references from the Archival Legacy Investigations of Circumstellar Environments (ALICE) project\footnote{\url{https://archive.stsci.edu/prepds/alice/}} (PI: R.~Soummer; \citealp{choquet14, hagan18}). We reduced the data with the non-negative matrix factorization method \citep[NMF:][]{ren18} using $30\%$ of the most correlated references with $50$ sequentially constructed NMF components. To recover the true surface brightness of these disks, we adopted a forward modeling approach assuming simple geometric models for debris architecture \citep{augereau99} and analytical SPFs \citep[e.g.,][]{{hg41}}. Due to the high computational cost of NMF component calculation \citep{ren18}, we saved the components computed in data reduction for subsequent forward modeling. We present the reduced images in Fig.~\ref{fig-nicmos-110} and Fig.~\ref{fig-nicmos-160} for filters F110W and F160W, respectively.

As opposed to the classical PSF subtraction method used for the STIS observations where there are dedicated stable reference star images, the NMF algorithm used for NICMOS -- which was shown to be able to better extract faint signals with higher quality than previous methods \citep[e.g.,][]{ren18, ren21} -- still introduces certain levels of overfit of disk signals. This is due to the diversity in stellar types, instrument observing conditions, and image stability in archival NICMOS observations, which makes the reference images not able to fully capture target PSFs for all observations in the near-infrared. To recover the surface brightness of a NICMOS disk, we did not adopt the scaling factor in \citet{ren18} that requires stable PSFs. Instead, we estimated the throughput of the algorithm by performing forward modeling to capture the PSF variation in the NICMOS archive. Specifically, we adopted the \citet{millarblanchaer15} code to create a disk model whose dust particles follow analytical SPFs in \citet{hg41}, and modified them in our study. 

To depict the spatial geometry of a debris disk, we used the \citet{ren21} modification of the \citet{millarblanchaer15} code: a combined power law in the disk mid-plane, and a vertical Gaussian dispersion, see \citet{augereau99}. In cylindrical coordinates, the disk follows

\begin{equation}\label{eq-powerlaw}
\rho(r, z)\propto \left[\left(\frac{r}{r_{\rm c}}\right)^{-2\alpha_{\rm in}}+\left(\frac{r}{r_{\rm c}}\right)^{-2\alpha_{\rm out}} \right]^{-\frac{1}{2}} \exp\left[-\left(\frac{z}{hr}\right)^2 \right],
\end{equation}
where $r_{\rm c}$ is the critical radius, $\alpha_{\rm in} > 0$ and $\alpha_{\rm out}<0$ are the asymptotic power law indices when $r \ll r_{\rm c}$ and $r \gg r_{\rm c}$, respectively. Although the scale height parameter is $h=0.04$ from a theoretical study by \citet{thebault09}, we note that edge-on disks may deviate from this value and thus retrieve it in our disk modeling procedure. To account for the inner and outer clearing radii beyond which there is no dust particles, $r_{\rm in}$ and $r_{\rm out}$, we only evaluate Equation~\eqref{eq-powerlaw} when $r_{\rm in} < r  < r_{\rm out}$, and it equals 0 otherwise. To depict the SPF of the scatterers in a debris disk, we adopted a two-component Henyey--Greenstein function \citep[e.g.,][]{chen20}, since the original analytical phase function in \citet{hg41} is monotonous, however that monotonicity is not always observed in actual debris disk observations \citep[e.g.,][]{stark14, chen20}. 

For each target, we first generated a model disk image, then convolved it with the corresponding NICMOS point source PSF created by {\tt TinyTim} \citep{tinytim}\footnote{\url{http://tinytim.stsci.edu}} using the effective temperature of the star from Table~\ref{tab:prop}. We subtracted the convolved disk from the observations to reperform NMF reduction using the originally calculated NMF components to reduce computational cost. In fact, for the debris disks in this study, we did not see major differences on re-computing the NMF components: this is likely due to the fact that the PSF wings are sufficiently brighter than debris disks in the data analyzed here, thus the latter do not contribute significantly to the selection of best-matching reference images. In comparison, when circumstellar disks are relatively brighter than PSF wings, we do indeed expect improvement of data reduction quality with NMF component re-computation (e.g., HD~100453 with VLT/SPHERE: \citealp{xie23}).

We distributed the calculation and forward modeling process using {\tt DebrisDiskFM} \citep{ren19} on a computer cluster, and explored the parameter space with {\tt emcee} \citep{emcee}. The best-fit models minimize the residuals by maximizing the log-likelihood, 

\begin{align*}\label{eq-loglike}
\ln\mathcal{L}\left(\bm{\Theta}\mid X_{\rm obs}\right) = &-\frac{1}{2}\sum_{i=1}^{N}\left(\frac{X_{{\rm obs}, i} - X_{{\rm model}, i}}{\sigma_{{\rm obs}, i}}\right)^2\\
	&- \sum_{i=1}^{N}\ln\sigma_{{\rm obs}, i} - \frac{N}{2} \ln(2\pi), \numberthis
\end{align*}
where we have assumed that the pixels $i$ follow independent normal distributions, with $X_{\rm obs}$ and $X_{\rm model}$ denoting the observation and model datasets, respectively. To quantify the uncertainty, we first obtained the algorithmic throughput of the best-fit model by comparing the model with the NMF reduction, then performed uncertainty measurement on the original individual NMF reductions with throughput correction.

\subsection{Data for joint analysis}

Given that the observed debris disks are of different inclinations, and that scatterers in debris disk systems redistribute incident light to different directions with varying intensity via SPFs \citep[e.g.,][]{stark14, milli17}, we measured the light with ${\approx}90^\circ$ scattering angle to minimize such effects to enable a uniform comparison of different systems. We used the regions annotated in Appendix~\ref{sec-app-color-locations} for measurements on the signal and background for both instruments. Specifically for NICMOS, by comparing our reduction of the original dataset with the best-fit convolved disk model, we can quantify the algorithmic throughput from the NMF post-processing procedure by dividing the NMF-reduced data with the best-fit model. We performed photometry on originally reduced data, subtracted flat halo backgrounds, and corrected the throughput measured from forward modeling. By doing so rather than performing measurements on the best-fit models, we expect to better capture the minor variations in observed disk signals. 

We obtained the regions for birth ring photometry and halo background measurements as follows. Using the HD~181327 system as an example, we first identified the debris birth ring in Figure~\ref{fig-region-and-background}(q) using the ring parameters (e.g., semi-major axis, position angle, inclination) from \citet{stark14}, then calculated for each pixel its scattering angle and associated angle uncertainty assuming an infinitely thin disk following \citet[][Appendix A therein]{ren19}. To identify the pixels that host birth ring signals, if a pixel's $1\sigma$ range of scattering angles overlaps with the $[80^\circ, 100^\circ]$ interval, we categorize it as a birth ring pixel with ${\approx}90^\circ$ scattering angle. To reduce certain contribution from the halo signals, we chose the pixels that are $1.5$ times the distance of the birth ring from the star for HD~181327, and calculated their mean for a flat halo background removal in further steps. To further assess the variation of halo background at different locations surrounding HD~181327, we measured the halo background at distinct locations with varying region area (while avoiding known birth ring signals), and we observed no significant difference from the measured trends in Sect.~\ref{sec-ana}.

For all debris systems, as a result, removing flat halo backgrounds induced minor deviation on the birth ring signals regardless of the location of the background pixels, since halos can be one or two orders of magnitude fainter than the birth rings \citep[e.g.,][]{schneider14, schneider18, ren19, ren21}. In fact, the detected STIS halos in Fig.~\ref{fig-stis} and in NICMOS images are only evident in log scale display. Halo background removal in linear scale, as well as the variation of halo signals within the chosen regions, has minor influence (${<}0.5\sigma$) on the extracted birth ring signals or the trend of birth ring color in Sect.~\ref{sec-ana}.

We also note that for nearly edge-on systems (e.g., AU~Mic, Beta~Pic, HD~32297, HD~141943) where we performed measurements on the ansae of the debris disks, the measurements can actually probe a range of scattering angles that can deviate significantly from ${\approx}90^\circ$. To explore possible measurement biases for these targets, as well as the impact of internal halo flat background at different regions for all targets, we varied the areas of regions for analysis for all systems by increasing or decreasing the signal and background extraction areas in Appendix~\ref{sec-app-color-locations} by factors of up to 4 either individually or jointly, and we did not observe statistically significant changes in our results or their interpretation in this study. We therefore adopt the regions identified in Appendix~\ref{sec-app-color-locations} for further analysis on both birth ring color and flat halo background removal.

\section{Analysis}\label{sec-ana}
We computed the STIS$-$NICMOS color of the disk images as follows. We first computed the reflectance in different filters for each system, then obtained the color for them. 
\subsection{Reflectance}
We obtained the instrument response\footnote{The instrument response in this study refers to stellar flux density integrated in the instrument filters unless otherwise specified.} for the stars in units of Jy by calculating the unobstructed instrumental response to the \citet{kurucz93} star models using {\tt pysynphot}  \citep{pysynphot}, where the inputs are their effective temperature ($T_{\rm eff}$), $V$-band magnitude, and surface gravity ($\log g$) in Table~\ref{tab:prop}. For NICMOS F110W or NICMOS F160W, the parameter is {\tt `nicmos,2,f110w'} or {\tt `nicmos,2,f160w'}, respectively. For STIS, {\tt `stis,ccd,a2d4'}.\footnote{\url{https://pysynphot.readthedocs.io/en/latest/appendixb.html}} We summarized the instrument response of the two coronagraphs in Table~\ref{tab:expinfo}. 

For the pre- and post-NCS eras of NICMOS operation (i.e., Era 1 and Era 2, respectively; see, e.g., \citealp{schultz03}) where the sensitivities of the instrument are distinct, we adopted different {\tt PHOTFNU} values to convert instrument counts to physical units of Jansky. For the two eras, the {\tt PHOTFNU} parameter for F110W is $1.84724\times10^{-6}$ and $2.03470\times10^{-6}$, respectively. For F160W, $1.21121\times10^{-6}$ and $1.49585\times10^{-6}$. The {\tt pysynphot} values correspond to Era 2 observations, for Era 1 observations we thus first multiplied an instrument count rate by the {\tt PHOTFNU} value in Era 2, then divided it by the {\tt PHOTFNU} value in Era 1, to obtain the count rate in Era 1. For each image, we obtained the fraction of light reflected by the debris disk via dividing the calibrated image by the {\tt pysynphot} rates for the star. We used these fraction images for color analysis.

\subsection{Dust Color}
To obtain the STIS$-$NICMOS color for a disk, we averaged $2\times2$ NICMOS pixels -- 1 NICMOS pixel is $75.65$~mas -- into 1 bin, and $3\times3$ STIS pixels -- 1 STIS pixel is $50.72$~mas -- into 1 bin, with each bin being a square with approximately $150$~mas in length. We then divided the binned STIS image by the square of the ratio between the width of the  STIS bin ($152.16$~mas) and that of the NICMOS bin ($153.3$~mas) to account for spatial scale difference, and converted the fraction values to magnitudes. To compare the reflectance in the two wavelengths for dust color, we subtracted the NICMOS magnitude from the STIS magnitude. In this way, a positive STIS-NICMOS value means the disk relatively scatters more light in NICMOS than in STIS, i.e., ``red scatterer'', while taking into account of the effect in the intrinsic brightness of the host star at different wavelengths.

We computed the ansae color along the major axes of the disks, i.e., a scattering phase angle of ${\approx}90^\circ$ between the incident light and the reflected light rays, to minimize the dependence of scattering intensity as a function of scattering phase angles \citep[i.e., SPF, e.g.,][]{hedman15}. See Sect.~\ref{sec-color-general} for a discussion on the contributions of signals from unbound particles (i.e., ``flat halo background''), and the regions used for their removal. See Appendix~\ref{sec-app-color-locations} for the regions used for color extraction and background removal.

\begin{figure}[htb!]
	\includegraphics[width=0.46\textwidth]{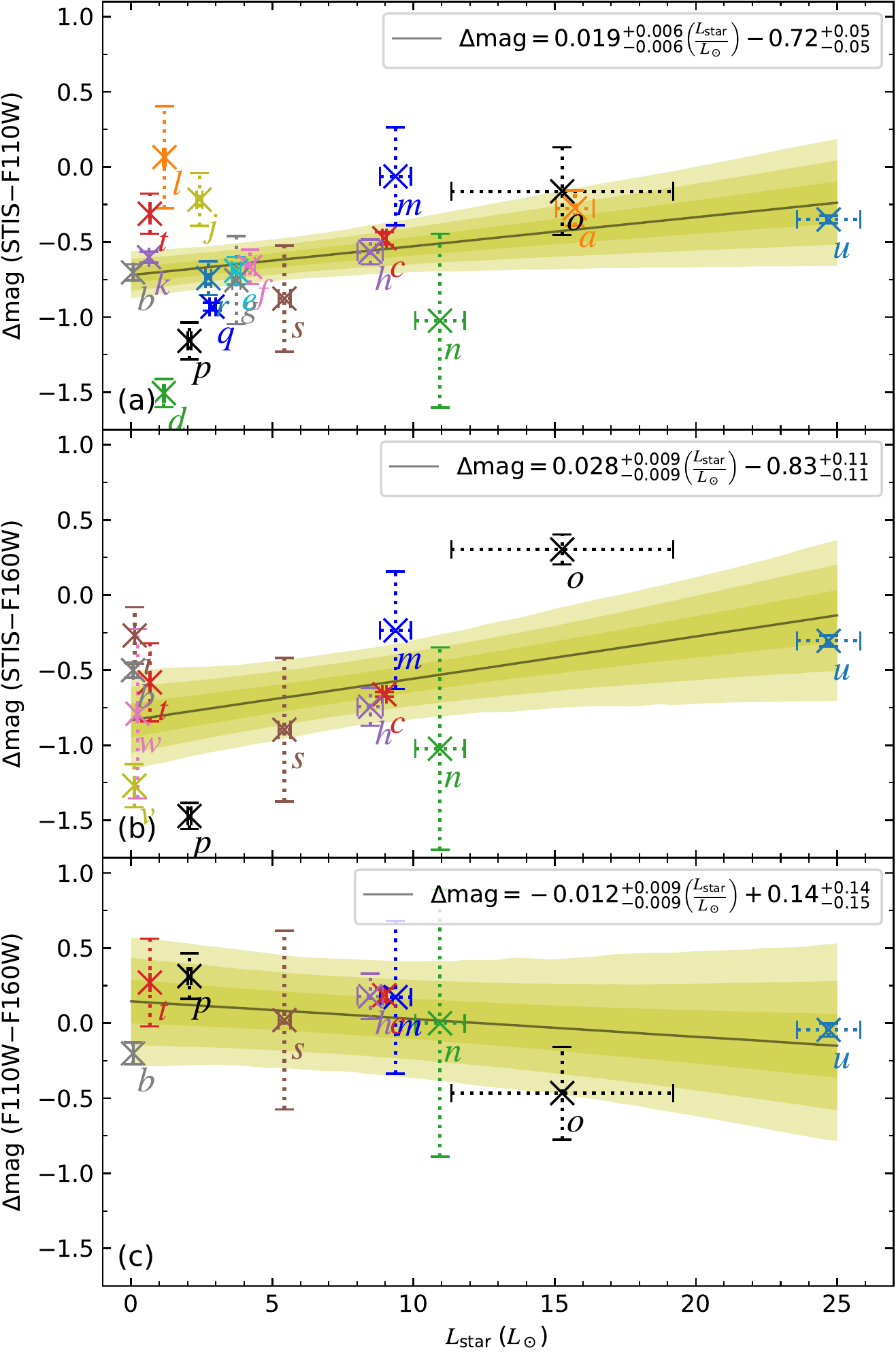}
    \caption{Dust color at $80^\circ$--$100^\circ$ scattering angle as a function of stellar luminosity. The letters next to the colorbars are the letter identifiers for the targets in Table~\ref{tab:prop}. Panels (a) and (b) suggest that dust particles scatter light more efficiently at shorter wavelengths for less luminous stars. Although the trend in panel (c) does not agree with the other two, it is marginal and likely impacted by smaller sample size and data reduction artifacts. The shaded areas are $1\sigma$, $2\sigma$, and $3\sigma$ confidence bands from bootstrapping fit.  See Table~\ref{tab:data} for the color values.}
    \label{fig-color-lum}    
\end{figure}

We present the dust color at the ansae of the birth rings as a function of stellar luminosity, obtained from \citet{tess_input_catalog}, in Fig.~\ref{fig-color-lum}. Comparing STIS with NICMOS observations, we notice that the general color is blue, while it becomes more neutral when stellar luminosity increases. In comparison with existing debris disk color studies comparing STIS and NICMOS \citep[e.g.,][]{ren19, ren21}, the observed colors are consistent within $2\sigma$ despite different color extraction methods. Nevertheless, for HR~4796A in STIS and F110W, although \citet{debes08} and \citet{rodigas15} obtained red colors  for the entire disk and the ansae, respectively, their results could have been compromised by the fact that certain signals were previously regarded as background before \citet[][Fig.~9 therein]{schneider18} and removed then. In fact, a blue ansae color measured for HR~4796A in this study is instead in agreement with the simulations from \citet{thebault19}, where the authors expected blue colors for debris birth rings for all A-type stars.

We observe that the F110W and F160W observations have a nearly neutral color, as well as a marginal trend with stellar luminosity in Fig.~\ref{fig-color-lum}. A neutral color within the NICMOS wavelengths could rise from multiple aspects. First, the two NICMOS filters are adjacent to each other in wavelength in Fig.~\ref{fig-transmission}, which might not provide distinctive difference from dust properties. Second, the NICMOS data were observed under less stable instrument conditions than STIS: although the NMF data reduction and forward modeling steps had outperformed other classical or statistics-based methods in the results, the results are still dominated by instrument instability or incomplete reference image sampling in NICMOS observations. Third but not least, the sample size of debris disks observed in these filters are smaller than when they are compared with STIS observations. Due to these aspects, we do not further discuss the trustworthiness of the color results within NICMOS wavelengths or their implications here.

\subsection{Color-albedo distribution: $90^\circ$-scattering albedo}\label{sec-color-albedo}

In planetary science studies on Solar System minor objects (e.g., asteroids, comets, and zodiacal light), spectral gradient and albedo can show different properties of these objects. The normalized reflectivity gradient follows $S'=\frac{1}{\bar{S}}\frac{{\rm d}S}{{\rm d}\lambda}$, where $S$ is the reflectance at wavelength $\lambda$, and $\bar{S}$ is the average reflectance \citep[e.g.,][]{yang15}. Under this convention, positive $S'$ indicates that the scatterers are more efficient in scattering photons in longer wavelengths (defined as a ``red'' color in our study), see Figs.~1 and 6 in \citet{yang15} for a comparison between zodiacal light (red color in their Fig.~6) and different asteroids.

Noticing the fact that the measurement for an asteroid normally has a dominant scattering angle, while the resolved debris disks by \textit{HST} have a range of scattering angles depending on their inclinations, we calculated a location-specific albedo for the debris disk samples in this study. Our definition of albedo is performed on the resolved debris disk only for those regions that satisfy one criterion: within which the scattering angles of the dust particles are between $80^\circ$ and $100^\circ$.

\subsubsection{Albedo measurements}

In an observed disk image in Fig.~\ref{fig-stis}, the area of regions with $90^\circ\pm10^\circ$ scattering angles can occupy a fraction $f_{[\rm 80^\circ, 100^\circ]} \in (0, 1]$ of the entire disk in the disk plane depending on the inclination of the disk: for a face-on disk, the entirety of the disk image has a $90^\circ$ scattering angle; for an edge-on disk, only the on-sky ansae of the birth ring (rather than the entirety of the major axis) have ${\sim}90^\circ$ scattering angle. To correct for this inclination-induced effect on the total scattered light at ${\sim}90^\circ$ scattering angle, and recover all the scattered light that are not fully captured by the telescope due along our line of sight, our recovery of $90^\circ$-scattering albedo follows

\begin{align}
\alpha_{[\rm 80^\circ, 100^\circ]} &= \frac{\frac{F_{\rm disk}/f_{[\rm 80^\circ, 100^\circ]}}{F_{\rm star}}}{ \frac{F_{\rm disk}/f_{[\rm 80^\circ, 100^\circ]}}{F_{\rm star}} + \frac{L_{\rm IR}}{L_{\rm star}}}  \label{eq-90-albedo-0}\\ 
						&= \frac{\frac{F_{\rm disk}}{F_{\rm star}}}{ \frac{F_{\rm disk}}{F_{\rm star}} + \frac{L_{\rm IR}}{L_{\rm star}}\times f_{[\rm 80^\circ, 100^\circ]}}, \label{eq-90-albedo}
\end{align}
where the disk flux $F_{\rm disk}$ is integrated in the observed disk region that satisfies the regional criteria. Indeed, by recovering the entire region of scattered light in Equation~\eqref{eq-90-albedo-0} using the partially observed data via $\frac{F_{\rm disk}}{f_{[\rm 80^\circ, 100^\circ]}}$, it is equivalent to multiplying the infrared excess of the disk  $\frac{L_{\rm IR}}{L_{\rm star}}$ by the fraction of the disk region in Equation~\eqref{eq-90-albedo}.

\begin{figure}[htb!]
\centering
\includegraphics[width=0.5\textwidth]{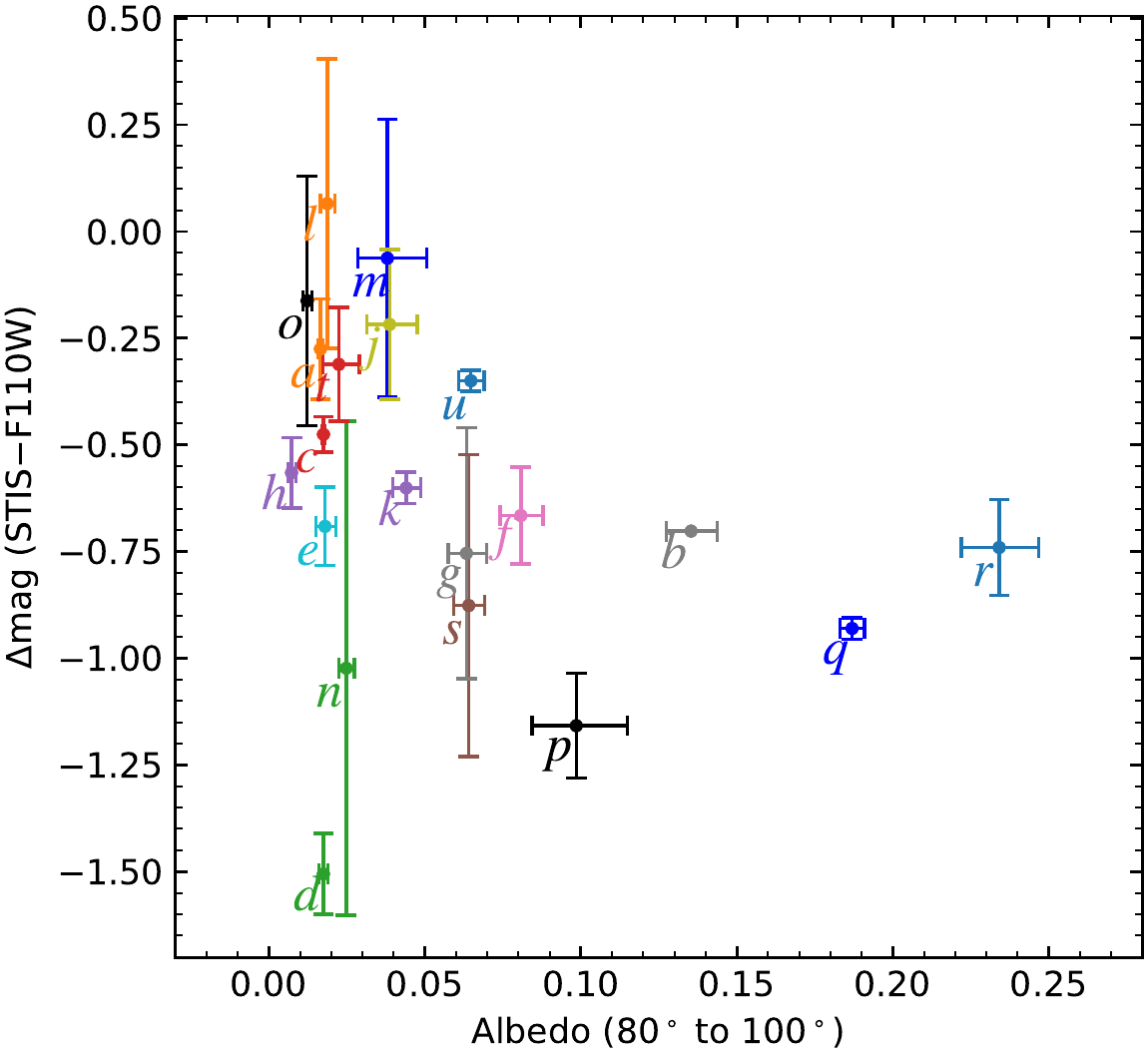}
    \caption{Disk color and $90^\circ$-scattering albedo distribution defined in Equation~\eqref{eq-90-albedo}. The likely L-shaped clustering of color-albedo distribution might resemble that of Solar System objects, which might indicate different formation history or composition of debris birth rings. See Table~\ref{tab:data} for the albedo values.}
    \label{fig-color-albedo}    
\end{figure}

Using the debris ring surface brightness values from STIS, we present the color--albedo measurements in Fig.~\ref{fig-color-albedo}. For the infrared excess values, we adopted the infrared excess data for HD~141569 from \citet{mawet17} and the rest from \citet{cotton16}. We notice a likely L-shaped clustering of debris disk albedo and color for the samples. In comparison with Solar System objects, only B-type and some of C-type asteroids, both of which are carbonaceous and belong to C-group asteroids \citep{tholen1989}, have blue color while other commonly observed S-type (siliceous) and X-type (metal-rich) asteroids are reddish \citep[e.g.,][]{yang15, mahlke2022}.

\subsubsection{Comparison with Solar System minor objects}
In visible to near-infrared wavelengths, most of the icy Kuiper Belt objects are red \citep[e.g.,][]{Tegler00, jewitt01, Delsanti06}. Nevertheless, $\rm EL_{61}$-group objects, which could have formed from a giant impact that removed ice mantles \citep[e.g.,][]{Barkume06, Brown07}, are slightly bluish \citep[][]{merlin2007, pinilla2008}. The $\rm EL_{61}$ spectrum can be explained by a large amount of crystalline and amorphous water ice on the surface \citep{Trujillo07, merlin2007}. The albedos of Kuiper Belt objects, which are not necessarily measured at ${\approx}90^\circ$ scattering angles as for debris disks in this study, have a large range from 0.01 to 0.8, with the majority of them below 0.2 \citep[e.g., Fig.~3 of][]{stansberry2008}. As for the Kuiper Belt dust, direct infrared observation is still not practical due to contamination of thermal emissions from the zodiacal cloud \citep{jewitt2008,brown2012}.

With an overarching caveat that the color--albedo studies on Solar System objects and on debris disks in Fig.~\ref{fig-color-albedo} are on objects with distinct sizes (${\sim}$km-sized objects and ${\sim}\mu$m-sized particles, respectively), the color-albedo distribution of the debris disks here might qualitatively resemble some C-group asteroids and a very few Kuiper Belt objects \citep[e.g.,][Fig.~1 therein]{yang15}. Nevertheless, given that the two albedos are calculated differently, although the debris disk albedos might resemble qualitatively most of C-type asteroids and Kuiper Belt objects, it does not suggest that the debris disk dust is made of materials that are identical to these Solar System objects.

In fact, in the young Solar System, dynamical processes can mix the minor objects within along the radial direction \citep{demeo2014}. As a result, the current spatial locations of Solar System minor objects does not match their initial locations. Therefore, the colors of the planetary objects can change over time from mechanisms such as space weathering, in which high energy particles from the Sun and cosmic ray bombard these minor objects \citep{hapke2001} with a timescale shorter than 1 Myr \citep{vernazza2009, demeo2023}. Space weathering could cause both the reddening \citep{binzel2001} and the bluing \citep{moroz2004} of minor objects, depending on the size of the grains \citep{thompson2020} and the composition \citep[e.g. C-type asteroids become bluer while S-type ones become redder:][]{nesvorny2005}.  The color change mechanisms for Solar System minor objects could further complicate the implications for the ensemble properties for the measurements of debris disks in this study.

Among Solar System minor objects, Q-type asteroids are considered to have fresh surfaces -- which could retain pristine materials that might resemble debris disk dust -- and are composed of ordinary chondrites. However, \citet{hasegawa2019} showed that the color of minor object spectra could be the consequence of space weathering on grains larger than 100 microns. Therefore, given that the measured colors of Solar System objects are likely on dust that are ${\sim}100~\mu$m while the typical sizes of debris disk dust are ${\sim}1~\mu$m in scattered light here, there might not exist pristine materials on the surface of the current Solar System objects, and thus a direct comparison of the colors between debris disks and Solar System objects is not feasible. Although we cannot directly match debris disks with Solar System minor objects, the likely L-shaped color-albedo distribution of the debris disks in Fig.~\ref{fig-color-albedo} might indicate not only the difference in dust composition, but also different (levels of) activities such as space weathering in the observed debris systems.

\subsubsection{Debris disk color-albedo clustering}

The likely L-shaped clustering of debris disk albedo and color in Fig.~\ref{fig-color-albedo}, in comparison with that of Solar System objects albeit with a caveat in the different definition of albedos, indicates that the dust particles in different debris disk systems are formed differently and/or have different composition. Nevertheless, there exists an extra source of physically-motivated uncertainty for our measurements: the collisional simulation study by \citet{thebault19} suggested that the halo outside the birth ring could contribute to ${\sim}50\%$ of the flux up to ${\sim}50~\mu$m.

The contribution from halo grains can impact infrared excess measurements \citep{thebault19}, and consequently would bias the albedo values measured here: we assumed all the infrared excess -- which is a combination from the birth ring and the halo -- are from the debris ring in Equation~\eqref{eq-90-albedo-0}. Given that the infrared excess signal from the birth ring alone is not easily separable in the \citet{thebault19} study, we also investigated a possible lower limit of that signal. Specifically, to explore the influence of halo grains on SEDs, we adopted the infrared excess of cold belts from \citet{chen14} in which the authors performed two-belt fits to the SEDs. By applying the \citet{chen14} cold belt results to Equation~\eqref{eq-90-albedo-0}, with a caveat the actual infrared excess could be lower \citep[e.g.,][]{thebault19}, we only observed  quantitative offsets for Fig.~\ref{fig-color-albedo}, and the different clusterings of color-albedo did not change qualitatively. However, noticing that two-belt SED fits still cannot intrinsically separate the contributions from the birth ring and the halo properly, the significant infrared excess contribution from halo grains in the \citet{thebault19} study suggests that the actual albedo values should be different than those presented in Fig.~\ref{fig-color-albedo}.

\section{Discussion}\label{sec-diss}

\subsection{Blue color of debris disks}\label{sec-color-general}
Comparing STIS and NICMOS observations of the debris disks in scattered light in ${\sim}0.6~\mu$m and in ${\sim}1.1~\mu$m or ${\sim}1.6~\mu$m, we obtained a predominantly blue color at the ansae (${\approx}90^\circ$ scattering angle) of debris birth rings. The observed blue color can suggest the ubiquitous existence of the sub-micron-sized particles that scatter light more efficiently in shorter wavelengths than larger particles. Indeed, the theoretical simulation study in \citet{thebault19} did show that even for A-type stars that were previously expected to blow sub-micron-sized dust out, high fractional luminosity disks (${\gtrsim}10^{-3}$) surrounding them can still harbor a sufficient number of these unbound dust (Fig.~2 therein) that are enough to make debris disks appear blue in scattered light (Fig.~13 therein).

While we have identified certain halo background areas that could aid in reducing the impact of unbound particles, the predominantly blue color of debris disk birth rings suggests that sub-micron-sized particles are widespread in all the systems studied here. What is more, the observed predominantly blue color suggests that a simple flat background removal adopted here has limited impacts on removing unbound dust contribution, since contributions from the SPFs of unbound particles are not negligible, especially when there exists an enough number of such particles as in \citet{thebault19}. Moreover, more importantly, there are other factors that could make the removal of a flat halo background less practical. First, \citet{lee16} simulations has shown that the existence of eccentric planet(s) can perturb the surface density distribution of halo particles. Second, the regions that we used to measure halo background have different stellocentric distances from that of the birth ring, requiring a distance-based illumination correction. Third but not least, the number density and surface density distributions of the unbound grains in the halo is not identical to that in the birth ring, calling for more investigations on the simulation results in or beyond \citet{thebault19}. Together with these limitations in removing the contribution from unbound particles on disk color measurement, a measurement of unbound particles in debris halo is not necessarily representative of the contribution of them at other locations including the birth ring. After all, the majority of debris disk birth rings indeed have blue colors, especially since sub-micron-sized particles -- some of which are unbound when hosted by early-type stars -- naturally reside in birth rings \cite[e.g.,][]{thebault19}.

The blue debris ring color is more neutral for more luminous stars in Fig.~\ref{fig-color-lum}. For the more luminous early-type stars, there could exist a small but sufficient number of sub-micron-sized particles that are unbound to make the debris rings blue \citep{thebault19}. In comparison, the less luminous later-type stars can indeed retain sub-micron-sized particles \citep[e.g.,][]{arnold19}, and these bound particles could make debris rings appear blue. As a result, for debris disks orbiting stars with increasing stellar luminosity in Fig.~\ref{fig-color-lum}, the sub-micron-sized particles within can turn from bound to unbound (i.e., from M-type to A-type stars), making the disks more neutral. Although it is not feasible to completely remove the color contribution from unbound particles in this study, the general trend of the color being more neutral for more luminous stars, if true, could be in line with the expectation that less sub-micron-sized particles are bound for earlier type stars, see Sect.~\ref{sec-color-amin} for a correlation between color and expected blowout size of dust particles.

Moving forward, to observationally better reveal the debris ring color for bound particles in debris systems (e.g., Fig.~13 of \citealp{thebault19}, in which debris disks orbiting A-type stars turn from red to blue when unbound particles are taken into account), radiative transfer modelings of the SPFs for unbound particles are necessary to remove SPF effects at different scattering angles. Such modeling works would be achieved in principle by fitting the observed halo intensity as a function of scattering angle, then by extrapolating the brightness for the unbound particles at debris birth ring regions. However, these modelings could be challenging in terms of dust morphological model, size, composition, and computational feasibility \citep[e.g.,][]{tazaki18, tazaki19, arnold19}, since they should be more realistic in resembling interplanetary dust particles (IDPs), which are produced from asteroid or comets from the inner main Asteroid Belt to the Kuiper Belt and they do have aggregate or fractal morphology \citep[e.g.,][]{bradley03}. In addition, the  \citet{lee16} simulations showed that the existence of hidden planetary perturber(s) on eccentric orbits can change the surface density distribution of particles in both the birth ring and the halo. We leave such an analysis on extracting debris birth ring colors only for bound particles for future studies.

\subsection{Dust blowout size and disk color}\label{sec-color-amin}

A dust particle experiences the force balance between radiational pressure and gravity pull, and it becomes unbound when the former exceeds the latter on an orbital timescale. For a given stellar system, we can calculate the dust blowout size for non-porous dust using Equation~(5) of \citet{arnold19} while assuming the average radiation-pressure efficiency over the stellar spectrum to be unity for compact spheres. Substituting the values for spherical amorphous olivine particles which have a mass density of $3.3$~g~cm$^{-3}$ as in \citet{chen14}, we obtain

\begin{equation}\label{eq-abo}
a_{\rm BO} = 0.35~\mu{\rm m} \times \frac{L_{\rm star}}{M_{\rm star}},
\end{equation}
where $L_{\rm star}$ and $M_{\rm star}$ are in Solar units. See Column (10) of Table~\ref{tab:prop} for the corresponding dust blowout size. We note that we ignored dependences on dust properties such as composition and porosity \citep[e.g.,][]{arnold19} to obtain a systematic view of the size information.

We fit the measured dust color with blow-out size using a linear relationship. There are positive correlations between the STIS$-$NICMOS colors and dust blow out size, or $\Delta{\rm mag}_{\rm STIS-F110W} = 0.12_{-0.05}^{+0.05}a_{\rm BO} -0.74_{-0.06}^{+0.06}$ and $\Delta{\rm mag}_{\rm STIS-F160W} = 0.17_{-0.07}^{+0.07}a_{\rm BO} -0.85_{-0.12}^{+0.12}$, respectively. The positive correlations indicate that that larger dust scatters light relatively more efficiently at longer wavelengths, and thus make a debris system relatively redder. 

Although there is a negative relationship between F110$-$F160W color and dust blow out size, or  $\Delta{\rm mag}_{\rm F110W-F160W} = -0.07_{-0.08}^{+0.07}a_{\rm BO} -0.15_{-0.15}^{+0.16}$, the statistical significance is tangential. Given the facts that both F110W and F160W observations have undergone forward modeling procedure, and that only $10$ out of the $23$ systems have observations in both filters, such a negative relationship is likely impacted by data reduction artifact and small sample size. Nevertheless, we conclude that on the one hand, it is not fully valid to assume single composition or ignore porosity to calculate the actual blow out size for debris disk systems, and on the other hand, it is challenging to calculate dust color for adjacent filters when data reduction artifacts are non-negligible.

\subsection{Disk infrared excess and disk color}\label{sec-color-lirlstar}
\begin{figure}[htb!]
	\includegraphics[width=0.48\textwidth]{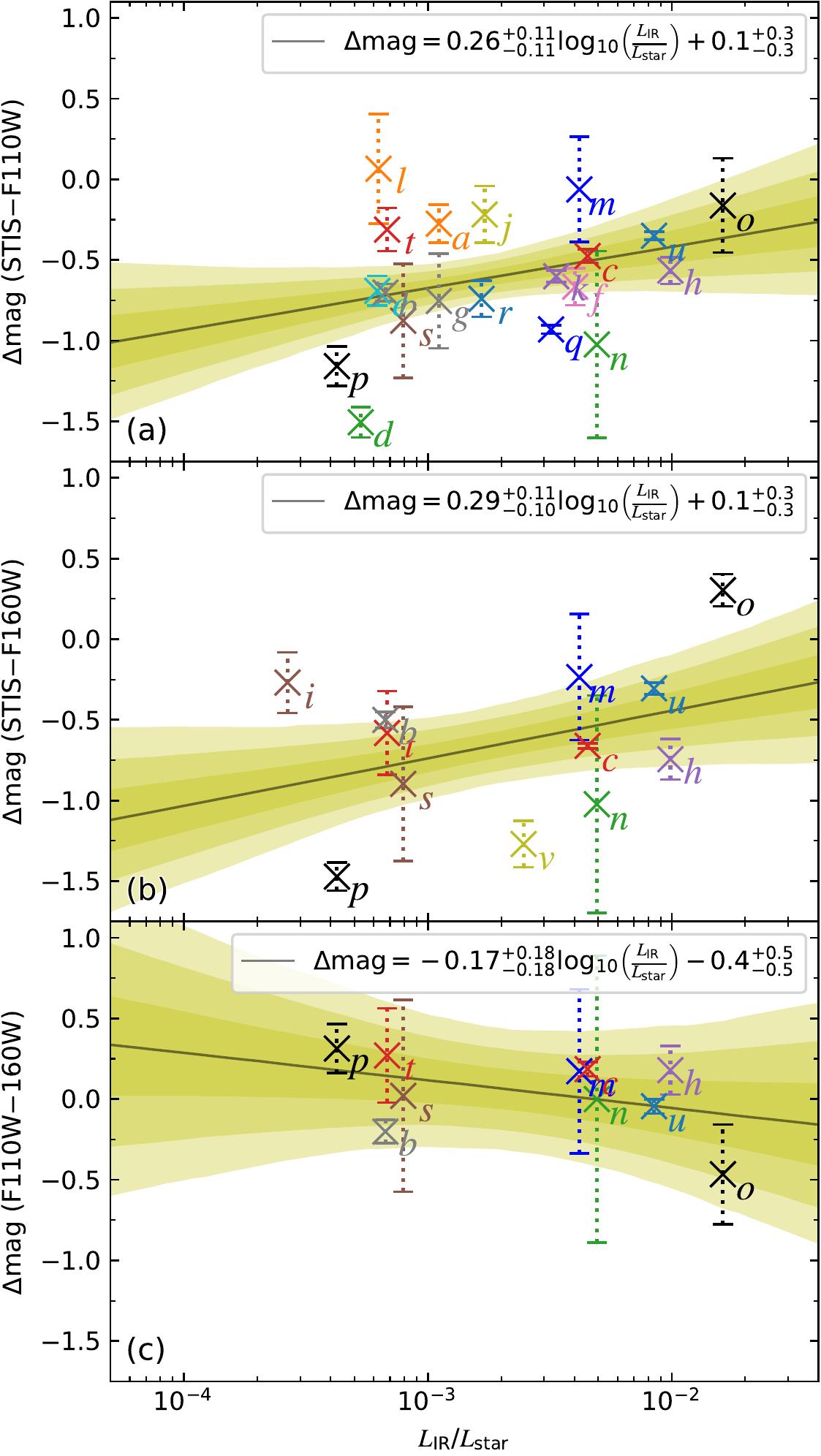}
    \caption{Dust color as a function of fractional infrared excess. Panels (a), (b), (c) are for STIS$-$F110W, STIS$-$F160W, and F110W$-$F160W, respectively. With a caveat that infrared excess and stellar luminosity are positively correlated in Sect.~\ref{sec-color-lirlstar}, there might exist marginal trends of the STIS$-$NICMOS color being more neutral for disks with higher fractional infrared excess.}
    \label{fig-color-lirlstar}    
\end{figure}

We present the disk color dependence on fractional infrared excess ($L_{\rm IR}/L_{\rm star}$) in Fig.~\ref{fig-color-lirlstar}, with the infrared excess data from Sect.~\ref{sec-color-albedo}. We observe a trend that disks with higher fractional infrared excess are more neutral in color. Following theoretical studies in which infrared excess decreases over time \citep[e.g.,][]{wyatt07, lohne08, gaspar13}, this color trend might be correlated with disk evolutionary stage, yet such a trend has a caveat that the debris disk colors are already under steady state in theoretical simulation studies \cite[e.g.,][]{thebault19}. What is more, more importantly, the fractional infrared excess is positively correlated with stellar luminosity in the samples in this study, making it probable that the former is not contributing to the color trend in Fig.~\ref{fig-color-lirlstar}.

To minimize the stellar luminosity contribution in the color dependence on fractional infrared excess, it is necessary to fit and remove stellar luminosity effects using Fig.~\ref{fig-color-lum}. However, with the high dispersion of the data in stellar luminosity for the systems in this study, as well as the fact that the samples in this study are not from a uniform survey, a proper removal of stellar luminosity influences for Fig.~\ref{fig-color-lirlstar} is beyond the scope of this study.

\section{Conclusion}\label{sec-sum}
By extracting the resolved debris disks using the \textit{Hubble Space Telescope} coronagraphs in visible and near-infrared light (${\sim}0.6~\mu$m, and ${\sim}1.1~\mu$m or ${\sim}1.6~\mu$m) using classical reference differential imaging and forward modeling in the STIS and NICMOS coronagraphs, we obtained the reflectance of these debris disks in scattered light. We observe that the color of these disks is predominantly blue, which suggests that the dust particles in these systems scatter shorter-wavelength light more efficiently than longer-wavelength light. 

In the albedo--color distribution of these systems, we notice the clustering of scatterers that could qualitatively resemble the clustering of Solar System objects, albeit with different definitions of albedos adopted. What is more, a qualitative resemblance does not indicate a compositional similarity. Were such a clustering true for debris disks, it could indicate different formation history and compositions for these debris systems.

The dust particles in these systems scatter relatively more efficiently (rather than absolutely more efficiently) in longer wavelengths, as the luminosity of the host star increases. This correlates with the expectation that more luminous stars can blow out relatively larger dust particles, and thus shifting the dust color towards the relatively redder direction. Nevertheless, given that there could still harbor a large amount of sub-micron-sized unbound particles even in A-type stars \citep{thebault19}, the measured blue color can rise from unbound particles in early-type stars as well as bound particles in late-type stars, which makes it challenging to separate these two kinds of contributions. A proper modeling and separation of scattered light contribution from unbound particles in the future is thus necessary to probe the birth ring color for bound particles in debris disks.

Accurate representation of the observed color and color-albedo distributions in debris disk systems requires the use of physically-motivated dust models. In comparison with existing attempts with Mie scatterers or a distribution of hollow spheres encountering difficulties in explaining observations \citep[e.g.,][]{milli19, ren19, chen20, arriaga20}, \citet{olofsson22} successfully reproduced the polarized light observations of HD~32297 with the \citet{tazaki18} models. The usage of more sophisticated or realistic models \citep[e.g.,][]{arnold19, tazaki19, TobonValencia22} or lab measurements \citep[e.g.,][]{munoz21} to resemble the IDPs that are originated from the Asteroid Belt and the Kuiper Belt, are necessary to properly depict the observed color as well as albedo for circumstellar disks in future works.

\begin{acknowledgements}
We thank the anonymous referee for their comments that increased the clarity, depth, and width of this paper. We thanks Xinyu Lu and Marco Delbo for helpful discussions. This work was funded by NASA through STScI Grant \# \textit{HST}-GO-15218.014-A for \textit{HST} GO-15218 program (PI: \'E.~Choquet). We are grateful for the productive discussions about dust scattering properties as part triggered by the EPOPEE (Etude des POussi\`eres Plan\'etaires Et Exoplan\'etaires) collaboration, supported by the French Planetology National Program (Programme National de Plan\'etologie, PNP) of CNRS/INSU co-funded by CNES. We thank in particular Jean-Charles Augereau for helpful discussions about minimum grain sizes in debris disks and J\'er\'emie Lasue for inspiring discussions about the diversity and properties of Solar System dust. E.C acknowledges funds from CNRS/PICS TACO-DESIRE program for supporting this research. This project has received funding from the European Research Council (ERC) under the European Union's Horizon 2020 research and innovation programme (PROTOPLANETS, grant agreement No. 101002188), and under the European Union's Horizon Europe research and innovation programme (ESCAPE, grant agreement No 101044152). Based on observations made with the NASA/ESA \textit{Hubble Space Telescope}, obtained from the data archive at the Space Telescope Science Institute. STScI is operated by the Association of Universities for Research in Astronomy, Inc.~under NASA contract NAS 5-26555. This research has made use of data reprocessed as part of the ALICE program, which was supported by NASA through grants \textit{HST}-AR-12652 (PI: R. Soummer), \textit{HST}-GO-11136 (PI: D.~Golimowski), \textit{HST}-GO-13855 (PI: \'E.~Choquet), \textit{HST}-GO-13331 (PI: L.~Pueyo), and STScI Director's Discretionary Research funds, and was conducted at STScI which is operated by AURA under NASA contract NAS5-26555. This research has made use of the SIMBAD database \citep{simbad}, operated at CDS, Strasbourg, France. This research has made use of the VizieR catalogue access tool, CDS,  Strasbourg, France (DOI:  \href{https://doi.org/10.26093/cds/vizier}{10.26093/cds/vizier}). The original description of the VizieR service was published in A\&AS 143, 23 \citep{ochsenbein00}. This research has made use of the SVO Filter Profile Service (\url{http://svo2.cab.inta-csic.es/theory/fps/}) supported from the Spanish MINECO through grant AYA2017-84089. The input images to ALICE processing are from the recalibrated NICMOS data products produced by the Legacy Archive project, ``A Legacy Archive PSF Library And Circumstellar Environments (LAPLACE) Investigation,'' (\textit{HST}-AR-11279, PI: G.~Schneider). This work has made use of data from the European Space Agency (ESA) mission {\it Gaia} (\url{https://www.cosmos.esa.int/gaia}), processed by the {\it Gaia} Data Processing and Analysis Consortium (DPAC, \url{https://www.cosmos.esa.int/web/gaia/dpac/consortium}). Funding for the DPAC has been provided by national institutions, in particular the institutions participating in the {\it Gaia} Multilateral Agreement. Part of the computations presented here were conducted in the Resnick High Performance Computing Center, a facility supported by Resnick Sustainability Institute at the California Institute of Technology. 
\end{acknowledgements}
\bibliography{refs}
\appendix
\section{Supplementary Materials}
\subsection{Exposure information of targets}\label{sec-app-expinfo}
We summarize the exposure time information for the targets in Table~\ref{tab:expinfo}.

\begin{table*}[htb!]
\setlength{\tabcolsep}{17pt}
\centering
\caption{Exposure Time and Instrument Response for Debris Disk Hosts\label{tab:expinfo}}
\begin{tabular}{c l| c c c| c c c}    \hline\hline
\multicolumn{2}{c|}{Instrument} 	& \multicolumn{1}{c}{F110W} &  \multicolumn{1}{c}{F160W} & \multicolumn{1}{c|}{STIS}  & \multicolumn{1}{c}{F110W} &  \multicolumn{1}{c}{F160W} & \multicolumn{1}{c}{STIS} \\ \hline
id & \multicolumn{1}{c|}{Target}	& \multicolumn{3}{c|}{Exposure Time (s)} & \multicolumn{3}{c}{Instrument Response (Jy)}\\ 
(1) & \multicolumn{1}{c|}{(2)} & (3) & (4) & \multicolumn{1}{c|}{(5)} & (6) &(7) &(8) \\ \hline
%\decimals
%\startdata
$a$ 	& 49~Ceti        	& 2335.57 	& ... 	&  1918.8 	&  11.0 	&  ... 	&  16.3 \\
$b$ 	& AU~Mic         	& 2687.53 	& 2687.53 	&  13050 	&  5.45 	&  7.65 	&  1.48 \\
$c$ 	& Beta~Pic       	& 127.78\tablenotemark{a} 	& 47.92\tablenotemark{a} 	&  2575.8 	&  85.7 	&  60.8 	&  83.3 \\
$d$ 	& HD~377         	& 4319.38 	& ... 	&  4596.6 	&  4.08 	&  ... 	&  2.76 \\
$e$ 	& HD~15115       	& 4607.27 	& ... 	&  11484.8 	&  6.21 	&  ... 	&  5.59 \\
$f$ 	& HD~15745       	& 1407.76 	& ... 	&  17730 	&  3.26 	&  ... 	&  2.96 \\
$g$ 	& HD~30447       	& 1343.76 	& ... 	&  4320 	&  2.41 	&  ... 	&  2.11 \\
$h$ 	& HD~32297       	& 1343.76 	& 1343.76 	&  13228.2 	&  1.29 	&  0.832 	&  1.58 \\
$i$ 	& HD~35650       	& ... 	& 1503.57 	&  6055.8 	&  ... 	&  4.11 	&  0.939 \\
$j$ 	& HD~35841       	& 1343.76 	& ... 	&  4350 	&  1.05 	&  ... 	&  0.816 \\
$k$ 	& HD~61005       	& 4607.34 	& ... 	&  13284 	&  2.65 	&  ... 	&  1.57 \\
$l$ 	& HD~104860      	& 5183.26 	& ... 	&  6360 	&  2.97 	&  ... 	&  2.05 \\
$m$ 	& HD~110058      	& 2303.6 	& 2303.6 	&  4708.8 	&  1.49 	&  0.956 	&  1.84 \\
$n$ 	& HD~131835      	& 2303.6 	& 2303.6 	&  4672.5 	&  1.57 	&  0.981 	&  2.04 \\
$o$ 	& HD~141569      	& 1215.76\tablenotemark{a} 	& 863.5\tablenotemark{a} 	&  16005.6 	&  2.98 	&  1.83 	&  4.03 \\
$p$ 	& HD~141943      	& 4607.34 	& 1535.56 	&  4479.6 	&  3.09 	&  2.74 	&  1.95 \\
$q$ 	& HD~181327      	& 1535.69 	& ... 	&  12812 	&  5.58 	&  ... 	&  4.51 \\
$r$ 	& HD~191089      	& 4607.34 	& ... 	&  7847.6 	&  4.89 	&  ... 	&  3.97 \\
$s$ 	& HD~192758      	& 3455.41 	& 3455.41 	&  4078.8 	&  4.49 	&  3.14 	&  4.48 \\
$t$ 	& HD~202917      	& 3519.51 	& 1407.74\tablenotemark{a} 	&  14034 	&  1.75 	&  1.62 	&  1.04 \\
$u$ 	& HR~4796A       	& 1951.65\tablenotemark{a} 	& 2367.57\tablenotemark{a} 	&  9525.6 	&  8.72 	&  5.05 	&  14.2 \\
$v$ 	& TWA~7          	& ... 	& 1215.76\tablenotemark{a} 	&  6718.8 	&  ... 	&  0.953 	&  0.177 \\
$w$ 	& TWA~25         	& ... 	& 1503.57 	&  14508 	&  ... 	&  0.755 	&  0.141 \\ \hline
\end{tabular}

\begin{flushleft}
\tiny  \textbf{Notes}: {Column (1): letter identifiers of targets. Column (2): target name. Columns (3), (4), and (5): exposure time using F110W, F160W, and STIS, respectively. Columns (6), (7), and (8): {\tt pysynphot} instrumental response for the unocculted stars using F110W, F160W, and STIS, respectively, with 3 significant digits.

$^a$Observed in NICMOS Era 1.}
\end{flushleft}
\end{table*}

\subsection{Color extraction locations}\label{sec-app-color-locations}
\begin{figure*}[htb]
\centering
	\includegraphics[width=\textwidth]{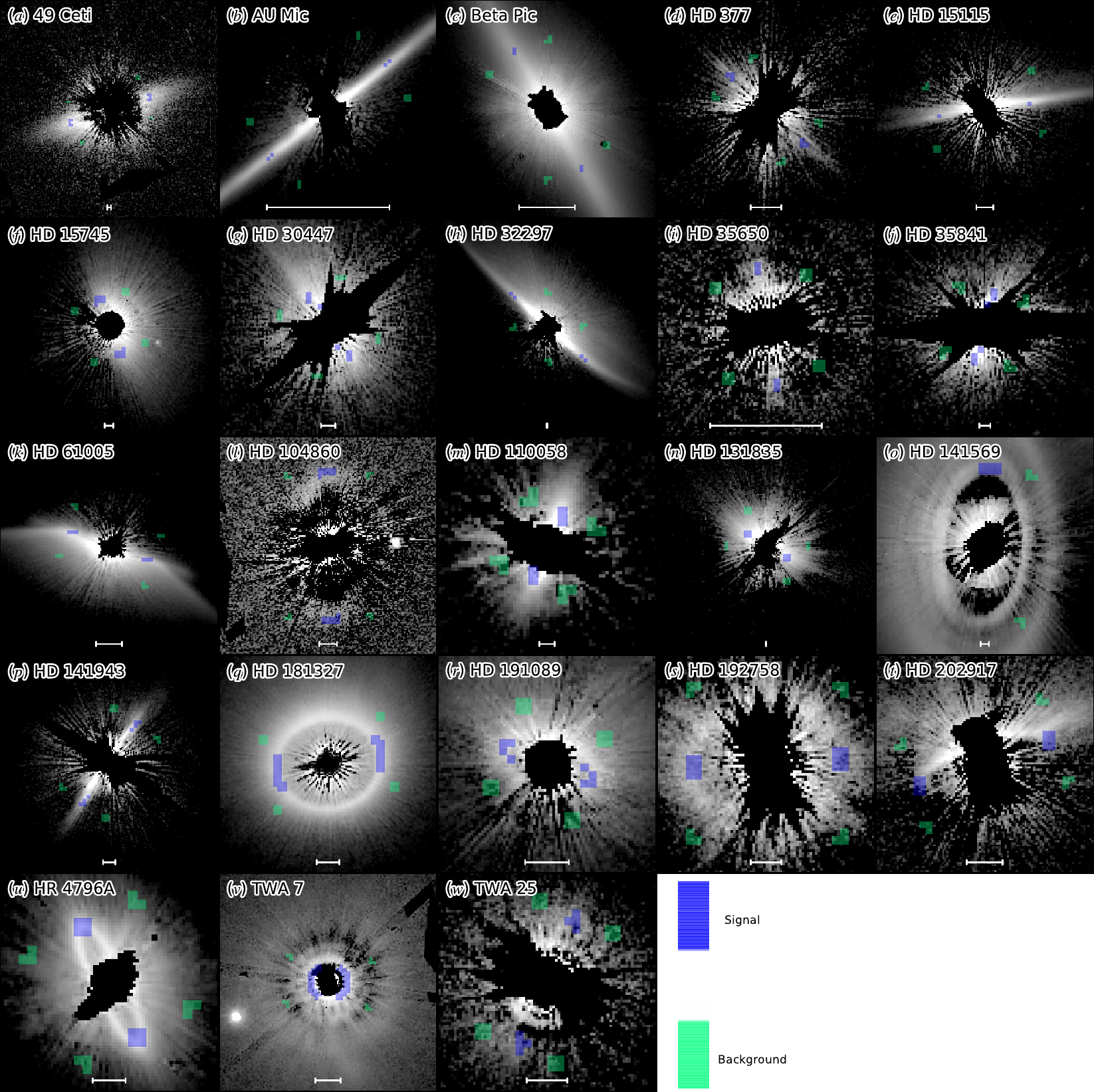}
    \caption{Regions used for dust color extraction (at $80^\circ$--$100^\circ$ scattering angle) and background removal overlayed onto STIS images. The surface brightness distributions of the disks are presented in log scale. When we changed the area of either or both regions by up to a factor of 4, the observed trends in both color and albedo did not have significant variation.
    }
    \label{fig-region-and-background}
\end{figure*}

We display the regions used to extract color information for the dust in Fig.~\ref{fig-region-and-background}.

\subsection{Color and albedo values}\label{sec-app-tab-color}
We present the extracted color and albedo values in Table~\ref{tab:data}.

\begin{table*}[htb]
\centering
\setlength{\tabcolsep}{17pt}
\caption{Color and albedo information measured in this study \label{tab:data}}
\begin{tabular}{c l c c c c }    \hline\hline
 id	&\multicolumn{1}{c}{Target} 	& STIS-F110W & STIS-F160W &  F110W-F160W & albedo(STIS-110W)\\
 (1)	& \multicolumn{1}{c}{(2)} & (3) & (4) &  (5)  & (6) \\ \hline
$a$	&49 Ceti	&$-0.28_{-0.12}^{+0.12}$	&$\cdots$		&$\cdots$	&$0.0165_{-0.0008}^{+0.0009}$\\ 
$b$	&AU Mic	&$-0.7_{-0.05}^{+0.05}$	&$-0.5_{-0.05}^{+0.05}$	&$-0.2_{-0.07}^{+0.07}$	&$0.135_{-0.008}^{+0.008}$\\ 
$c$	&Beta Pic	&$-0.48_{-0.04}^{+0.04}$	&$-0.661_{-0.015}^{+0.015}$	&$0.19_{-0.04}^{+0.04}$	&$0.0175_{-0.0004}^{+0.0004}$\\ 
$d$	&HD 377	&$-1.51_{-0.09}^{+0.09}$	&$\cdots$		&$\cdots$	&$0.0175_{-0.0014}^{+0.0015}$\\ 
$e$	&HD 15115	&$-0.69_{-0.09}^{+0.09}$	&$\cdots$		&$\cdots$	&$0.018_{-0.003}^{+0.003}$\\ 
$f$	&HD 15745	&$-0.67_{-0.11}^{+0.11}$	&$\cdots$		&$\cdots$	&$0.081_{-0.007}^{+0.007}$\\ 
$g$	&HD 30447	&$-0.8_{-0.3}^{+0.3}$	&$\cdots$		&$\cdots$	&$0.063_{-0.006}^{+0.006}$\\ 
$h$	&HD 32297	&$-0.57_{-0.08}^{+0.08}$	&$-0.74_{-0.13}^{+0.13}$	&$0.18_{-0.15}^{+0.15}$	&$0.0073_{-0.0011}^{+0.0013}$\\ 
$i$	&HD 35650	&$\cdots$		&$-0.27_{-0.19}^{+0.19}$	&$\cdots$	&$0.052_{-0.004}^{+0.005}$\\ 
$j$	&HD 35841	&$-0.22_{-0.18}^{+0.18}$	&$\cdots$		&$\cdots$	&$0.039_{-0.007}^{+0.009}$\\ 
$k$	&HD 61005	&$-0.6_{-0.04}^{+0.04}$	&$\cdots$		&$\cdots$	&$0.044_{-0.004}^{+0.005}$\\ 
$l$	&HD 104860	&$0.1_{-0.3}^{+0.3}$	&$\cdots$		&$\cdots$	&$0.019_{-0.002}^{+0.002}$\\ 
$m$	&HD 110058	&$-0.1_{-0.3}^{+0.3}$	&$-0.2_{-0.4}^{+0.4}$	&$0.2_{-0.5}^{+0.5}$	&$0.038_{-0.010}^{+0.013}$\\ 
$n$	&HD 131835	&$-1.0_{-0.6}^{+0.6}$	&$-1.0_{-0.7}^{+0.7}$	&$-0.0_{-0.9}^{+0.9}$	&$0.025_{-0.002}^{+0.003}$\\ 
$o$	&HD 141569	&$-0.2_{-0.3}^{+0.3}$	&$0.30_{-0.10}^{+0.10}$	&$-0.5_{-0.3}^{+0.3}$	&$0.0123_{-0.0014}^{+0.0016}$\\ 
$p$	&HD 141943	&$-1.16_{-0.12}^{+0.12}$	&$-1.47_{-0.09}^{+0.09}$	&$0.31_{-0.15}^{+0.15}$	&$0.099_{-0.014}^{+0.016}$\\ 
$q$	&HD 181327	&$-0.93_{-0.03}^{+0.03}$	&$\cdots$		&$\cdots$	&$0.187_{-0.004}^{+0.004}$\\ 
$r$	&HD 191089	&$-0.74_{-0.11}^{+0.11}$	&$\cdots$		&$\cdots$	&$0.234_{-0.012}^{+0.013}$\\ 
$s$	&HD 192758	&$-0.9_{-0.4}^{+0.4}$	&$-0.9_{-0.5}^{+0.5}$	&$0.0_{-0.6}^{+0.6}$	&$0.064_{-0.005}^{+0.005}$\\ 
$t$	&HD 202917	&$-0.31_{-0.13}^{+0.13}$	&$-0.6_{-0.3}^{+0.3}$	&$0.3_{-0.3}^{+0.3}$	&$0.022_{-0.005}^{+0.007}$\\ 
$u$	&HR 4796A	&$-0.35_{-0.02}^{+0.02}$	&$-0.31_{-0.04}^{+0.04}$	&$-0.04_{-0.04}^{+0.04}$	&$0.065_{-0.004}^{+0.004}$\\ 
$v$	&TWA 7	&$\cdots$		&$-1.27_{-0.14}^{+0.14}$	&$\cdots$	&$0.57_{-0.02}^{+0.02}$\\ 
$w$	&TWA 25	&$\cdots$		&$-0.8_{-0.6}^{+0.6}$	&$\cdots$	&$\cdots$	\\ \hline
\end{tabular}

\begin{flushleft}

{\tiny \textbf{Notes}: Column (1): letter identifiers of the targets in this paper. Column (2): target name. Column (3): STIS - F110W color. Column (4): STIS - F160W color. Column (5): F110W - F160W color. Column (6): $90^\circ$-scattering albedo for STIS-F110W color.}
\end{flushleft}

\end{table*}

\end{CJK*}
\end{document}